\documentclass[
  twocolumn,aps,prx,amsmath,longbibliography,
  amssymb,superscriptaddress,bibnotes
  ]{revtex4-1} 

\usepackage{hyperref}
\usepackage{graphicx}
\usepackage{mathtools}
\usepackage{hyperref}
\usepackage{color}
\usepackage{float}
\usepackage[export]{adjustbox}

\newcommand{\astcycl}{\mathrlap{\kern0.085em{\circlearrowright}}\ast}

\newcommand{\taucycl}{\mathrlap{\kern0.42em{\bullet}}\circlearrowright}

\begin{document}

\title{The anisotropic Harper-Hofstadter-Mott model: \\
 competition between condensation and magnetic fields}

\author{Dario H\"ugel}
\email{dario.huegel@physik.uni-muenchen.de}
\affiliation{Department of Physics, Arnold Sommerfeld Center for Theoretical Physics and Center for NanoScience, Ludwig-Maximilians-Universit\"at M\"unchen, Theresienstrasse 37, 80333 Munich, Germany} 

\author{Hugo U.~R.~Strand}
\affiliation{Department of Quantum Matter Physics, University of Geneva, 24 Quai Ernest-Ansermet, 1211 Geneva 4, Switzerland} 
\affiliation{Department of Physics, University of Fribourg, 1700 Fribourg, Switzerland} 

\author{Philipp Werner}
\affiliation{Department of Physics, University of Fribourg, 1700 Fribourg, Switzerland} 

\author{Lode Pollet}
\affiliation{Department of Physics, Arnold Sommerfeld Center for Theoretical Physics and Center for NanoScience, Ludwig-Maximilians-Universit\"at M\"unchen, Theresienstrasse 37, 80333 Munich, Germany} 
\date{\today} 


\begin{abstract}
We derive the reciprocal cluster mean-field method to study the strongly-interacting bosonic Harper-Hofstadter-Mott model. The system exhibits a rich phase diagram featuring band insulating, striped superfluid, and supersolid phases. Furthermore, for finite hopping anisotropy we observe gapless uncondensed liquid phases at integer fillings, which are analyzed by exact diagonalization. The liquid phases at fillings $\nu=1,3$ exhibit the same band fillings as the fermionic integer quantum Hall effect, while the phase at $\nu=2$ is $\mathcal{C}\mathcal{T}$-symmetric with zero charge response. We discuss how these phases become gapped on a quasi-one-dimensional cylinder, leading to a quantized Hall response, which we characterize by introducing a suitable measure for non-trivial many-body topological properties. Incompressible metastable states at fractional filling are also observed, indicating competing fractional quantum Hall phases. The combination of reciprocal cluster mean-field and exact diagonalization yields a promising method to analyze the properties of bosonic lattice systems with non-trivial unit cells in the thermodynamic limit.
\end{abstract}

\maketitle
\makeatletter
\let\toc@pre\relax
\let\toc@post\relax
\makeatother

%

\section{Introduction}

\begin{figure}[H]
  \includegraphics[scale=.9]{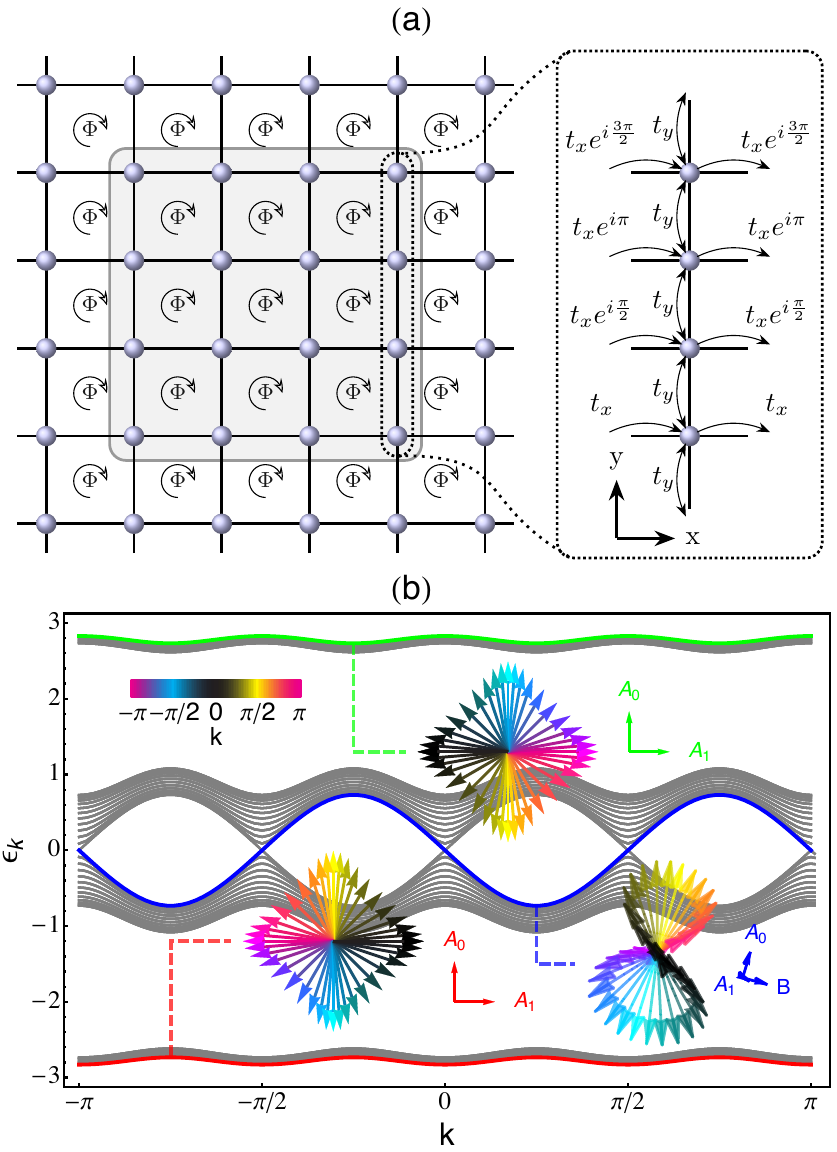}
  \caption{\label{fig:setup} Harper-Hofstadter model. (a) Setup of the single-particle hopping where each plaquette is pierced by a flux of $\Phi$. The $4\times 1$ unit cell for $\Phi=\pi/2$ is shown (dotted lines), where the arrows indicate the direction of the corresponding hopping processes. The $4 \times 4$ cluster employed in the RCMF approach is also shown (gray shaded area). (b) Single-particle dispersion for $\Phi=\pi/2$ and $t_x=t_y=1$. The precession of the $\hat{h}_{k,q}$ vector [Eq.\ (\ref{eq:Svec})] is shown for three states (red, blue and green) when varying $k$. The vector-colors indicate the values of $k$ (see colorbar).}
\end{figure}

Since the discovery of the quantum Hall effect \cite{Klitzing_QH,Tsui_QH,Laugh_QH}, the lattice geometry's influence on charged particles in magnetic fields has been the subject of extensive research. Prototypical models such as the non-interacting Harper-Hofstadter model (HHm) \cite{Harper_HHM,Hof_HHM} exhibit fractionalization of the Bloch bands with non-trivial topology (see Fig.\ \ref{fig:setup}),  manifesting in quantum (spin) Hall phases \cite{Bern_QSH,Gold_QSH,Hof}. Ultracold atomic gases with artificial magnetic fields \cite{Madi_ArtGa,Abo_Art,BlochRev,Lin_art,Dali_Art} enabled the experimental study of the non-interacting model \cite{Hof,Miyake_13,Atal_14,Aid_15}, while the effect of strong interactions on the band properties remains an open problem. While heating processes in the regime of strong interactions still represent a problem for cold atom experiments with artificial magnetic fields, recent experimental progress gives hope that this can be controlled in the near future \cite{Ket_Int,Grein_16}.

For bosons in the HHm with local interaction, i.e.\ the Harper-Hofstadter-Mott model (HHMm), previous theoretical studies  found fractional quantum Hall (fQH) phases, which have no counterpart in the continuum for strong fields~\cite{Mol_CFM}, using  exact diagonalization (ED) \cite{ED_05,ED_07,Mol_CFM}, composite fermion theory  \cite{Mol_CFM}, and the density matrix renormalization group (DMRG) on  a cylinder \cite{He_HHM}. Composite fermion studies also found evidence of a bosonic integer quantum Hall phase in bands with Chern number two \cite{Mol_C2}, also observed with ED \cite{Zeng_16} in the presence of next-neighbor hopping. In  a recent DMRG study \cite{He_HHM} a bosonic integer quantum Hall groundstate was also observed in the standard HHMm at filling $\nu=2$. However, the composite fermion approach is biased by the choice of the wavefunction \cite{Mol_CFM,Mol_C2}, while ED on small finite systems suffers from strong finite-size effects \cite{ED_05,ED_07,Zeng_16}.  

\pagebreak
The issue becomes especially challenging when going to strong fluxes, where extrapolation to the thermodynamic limit is impossible~\cite{Haf_EPL} and the overlap of the finite-size groundstate with Laughlin \cite{ED_05,ED_07} or composite fermion \cite{Mol_CFM} wavefunctions quickly decreases. DMRG, on the other hand, is restricted to cylinder geometries \cite{He_HHM,Pol_SPT,Motruk} and encounters huge numerical difficulties in the case of critical phases. Variational Gutzwiller mean field studies also found evidence of fQH phases \cite{Palm_08,Onur1,Onur2}, as well as striped vortex-lattice phases \cite{Palm_08}, but the variational basis is restricted by construction. 
The results of a recent cluster Gutzwiller mean field (CGMF) study~\cite{Natu_HHM} are likewise hard to interpret since the method breaks the translational invariance and the topology of the system.
 
To overcome these problems, we develop a reciprocal cluster mean field (RCMF) method, directly defined in the thermodynamic limit, which preserves the topology of the lattice, and yields excellent agreement  with numerically exact results for the Bose-Hubbard model. Further, we introduce an observable for the measure of topological properties in the presence of interactions.

We systematically map out the phase diagram of the strongly interacting HHMm as a function of the chemical potential and the hopping anisotropy. The phase diagram features band insulating, striped superfluid, and supersolid phases. At integer fillings we further observe highly anisotropic gapless uncondensed liquid phases, which are analyzed using exact diagonalization. For fractional filling, we find incompressible metastable states, indicating competing fQH phases. We define the respective order parameters, and present spatially resolved density, condensate-density, and current patterns.
Finally, we discuss how on an infinite cylinder with a single unit-cell in the $y$-direction the liquid phases become gapped and show a quantized Hall response to the adiabatic insertion of a magnetic flux. 

This paper is organized as follows. In Sec.\ \ref{Sec_Mod} the HHm and HHMm are discussed, and the method for measuring non-trivial topological properties is introduced. The RCMF method is derived and discussed in Sec.\ \ref{Sec_RCMF}, while the results for the HHMm are presented and discussed in Sec.\  \ref{Sec_Res}. Finally, Sec.\ \ref{Sec_Conc} is devoted to conclusions.

\section{Model}
\label{Sec_Mod}

\subsection{Harper-Hofstadter model}
\label{sec:nonint}

To facilitate the discussion for the strongly-interacting system, we first review the non-interacting HHm on the square lattice. The Hamiltonian is given by
\begin{equation}
  H_\Phi=
  - \sum_{x,y}\left(
    t_x e^{iy\Phi}b^{\dagger}_{x+1,y}b_{x,y}
  + t_y b^{\dagger}_{x,y+1}b_{x,y}
  \right)+\textit{h.c.}
  \label{eq:HofHarp_xy}
\end{equation}
with hopping amplitudes $t_{x/y}$ and annihilation (creation) operators $b^{\left(\dagger\right)}_{x,y}$. Each plaquette is pierced by a flux such that a phase $\Phi$ is picked up when going around it, as illustrated in Fig.\ \ref{fig:setup}a.
For $\Phi=2\pi/N_{\Phi}$ the unit cell can be chosen as  $N_{\Phi}$ sites in the $y$-direction.

Eq.\ (\ref{eq:HofHarp_xy}) is diagonalized by the transform $b_l(k,q)=\left(2\pi\right)^{-1}\sum_{x,j}e^{-i\left(kx+q\left(l+jN_{\Phi}\right)\right)}b_{x,l+jN_{\Phi}}$,
where $l\in\left[0,N_{\Phi}-1\right]$ and $k$($q$) are the momenta in $x$($y$)-direction. 
For even $N_{\Phi}$ the Hamiltonian reduces to $H_\Phi=\sum_{k,q}H_{k,q}$, with
\begin{align}
  H_{k,q}  = &
  -\sum_{l=0}^{N_{\phi}/2-1} 2t_x \cos\left(k-l\Phi\right)A_{l}(k,q) \nonumber \\ & - 2 t_y\cos(q) B(k,q) ,
  \label{eq:HofHarp_kq}
\end{align}
and
\begin{align}
  A_{l}(k,q) & = n_l(k,q)-n_{l+N_\phi/2}(k,q) \label{eq:Sz} \, , \\ 
  B(k,q) & =\frac{e^{-iq}}{2\cos(q)}\sum_{l}b^{\dagger}_{l+1}(k,q)b^{}_{l}(k,q)+\textit{h.c.},
  \label{eq:Sx}
\end{align}
For $\Phi=\pi/2$, used below, the system has three isolated topologically non-trivial bands, see Fig \ref{fig:setup}b. Here we use the notation of Ref.\ \onlinecite{Aid_15} where the central (super)band contains twice the number of states as compared to the two other bands. For a discussion of how the hopping anisotropy $t_{x}/t_y\neq 1$ affects the bandstructure, see Appendix \ref{hhm}.

The Hamiltonian $H_{\Phi}$ and Eq.\ (\ref{eq:HofHarp_kq}) can be rewritten in the compact notation
\begin{equation}
H_{\Phi}=\int dk dq \left(\vec{v}_{k,q}\cdot\vec{h}_{k,q}\right),\label{eq:h_phi}
\end{equation}
where $\vec{v}_{k,q}$ is a vector of scalars and $\vec{h}_{k,q}$ is a vector of operators
\begin{equation}
\vec{v}_{k,q}=\left(\begin{array}{l} -2t_x\cos\left(k\right) \\ -2t_x\cos\left(k-\frac{\pi}{2}\right) \\ -2t_y\cos\left(q\right)  \end{array}\right),\text{   }\vec{h}_{k,q}=\left(\begin{array}{l} A_0\left(k,q\right) \\ A_1\left(k,q\right)  \\ B\left(k,q\right)   \end{array}\right).\nonumber
\end{equation}

The operator $\vec{h}_{k,q}$ fully determines the momentum dependence of the non-interacting system, and we can apply the concept of parallel transport \cite{Qi_parallel}. The local Berry curvature at the point $(k,q)$ is proportional to the rotation of the unit-vector 
\begin{equation}
\hat{h}_{k,q}=\langle\vec{h}_{k,q}\rangle/|\langle \vec{h}_{k,q}\rangle|
\label{eq:Svec}
\end{equation}
 under an infinitesimal momentum shift. In fact, if $\hat{h}_{k,q}$ shows a non-trivial winding under transport on a closed path through the Brillouin zone, the Berry-curvature cannot be continuously deformed to a trivial one and the system is topologically non-trivial. The Chern number of the $n$th band is given by the number and direction of closed loops of $\hat{h}_{k,q}$, i.e.
\begin{equation}
c_n=\frac{\gamma_n}{2\pi},\nonumber
\end{equation}
where $\gamma_n$ is the solid angle subtended by $\hat{h}_{k,q}$ when taking the expectation value with respect to the single-particle eigenstates of the $n$th band and sweeping the momenta through the Brillouin zone. This is shown in Fig.\ \ref{fig:setup}b. For the lowest band $\langle A_{0}(k,0) \rangle$ and $\langle A_{1}(k,0)\rangle$ are shown while $\langle B(k,0)\rangle$ varies only slightly: $\hat{h}_{k,q}$ performs one anti-clockwise loop, corresponding to a Chern number of $c_0=-1$. Equivalently, for the central band $\hat{h}_{k,q}$ performs a double clockwise loop ($c_1=2$), while the highest band again has $c_2=-1$. 

The connection between the winding of $\hat{h}_{k,q}$ and the Hall conductivity can be seen from the example of the integer quantum Hall effect, i.e. the lowest band being completely filled with non-interacting fermions. Adding a magnetic flux $\Phi_{\rm y}$ piercing a torus of size $L_x\times L_y$ in $y$-direction can be achieved by transforming the hopping amplitudes as $t_x\rightarrow t_x e^{i\Phi_{\rm y}/L_x}$ for hopping processes in $+\hat{x}$, while taking the complex-conjugate in the opposite direction. The effect of this transform on the Hamiltonian (\ref{eq:h_phi}) amounts to 
\begin{equation}
\vec{v}_{k,q}\rightarrow \vec{v}_{k-\Phi_{\rm y}/L_x,q},\nonumber
\end{equation}
which is manifested  in a translation of the vector $\vec{h}_{k,q}$ with respect to the case without flux \textit{at each} momentum $(k,q)$, i.e.
\begin{align}
&\left\langle \Psi(\Phi_{\rm y})\left| \vec{h}_{k,q} \right|\Psi(\Phi_{\rm y})\right\rangle =\nonumber \\
&\left\langle \Psi(0)\left| \vec{h}_{k+\Phi_{\rm y}/L_x,q} \right|\Psi(0)\right\rangle,
\label{gauge}
\end{align}
where $ \left|\Psi(\Phi_{\rm y})\right\rangle$ is the many-body groundstate under the flux $\Phi_{\rm y}$.

The sole effect of $\Phi_{\rm y}$ is therefore a transform of the many-body groundstate such that at each momentum $(k,q)$ Eq.\ (\ref{gauge}) is fulfilled, resulting in a rotation of $\hat{h}_{k,q}$. Inserting a flux of $\Phi_{\rm y}=2\pi L_x/4$ yields the transform $A_0(k,q)\rightarrow A_1(k,q)$, $A_1(k,q)\rightarrow -A_0(k,q)$, and therefore $n_{l}(k,q)\rightarrow n_{l+1}(k,q)$. Adding a magnetic flux of $\Phi_{\rm y}=2\pi L_x/4$ is equivalent to translating the manybody groundstate by one site in the $y$-direction. 

If the lowest band is completely filled, the total number of particles on the torus is $L_x L_y/4$. Therefore adiabatically inserting a flux of $\Phi_{\rm y}=2\pi L_x/4$ results in $L_x L_y/4$ particles being translated by one site in $y$-direction, or equivalently a total number of $L_x/4$ particles being transported once around the periodic boundary in the $y$-direction. Consequently, adiabatically inserting a flux of $\Phi_{\rm y}=2\pi$ results in a quantized total transverse transport of a single particle around the periodic boundary.

\subsection{Harper-Hofstadter-Mott model}
\label{sec:interact}

We proceed with the study of the HHMm with interaction $U$, chemical potential $\mu$, and magnetic flux $\Phi=\pi/2$,
\begin{equation}
H=H_{\Phi}+\lim_{U \to \infty} \frac{U}{2}\sum_{x,y}n_{x,y}(n_{x,y}-1)-\mu\sum_{x,y}n_{x,y},
   \label{eq:HofHarp}
\end{equation}
in the hard-core limit $U\rightarrow\infty$. 

In contrast to the non-interacting case, for a finite interacting system the Berry curvature is defined with respect to boundary twisting angles \cite{Haf_EPL}, i.e.,
\begin{equation}
C=\frac{1}{2\pi}\int_0^{2\pi}d\theta_x\int_0^{2\pi}d\theta_y \left(\partial_{\theta_x}\mathcal{A}_{y}-\partial_{\theta_y}\mathcal{A}_{x}\right),
\label{eq:che}
\end{equation}
where $\mathcal{A}_{j}(\theta_x,\theta_y) =i\langle\Psi\left(\theta_x,\theta_y\right)|\partial_{\theta_{j}}|\Psi\left(\theta_x,\theta_y\right)\rangle$ is the Berry connection, $\Psi$ is the many-body groundstate, and $\theta_x$, and $\theta_y$ are twisting angles of the boundary conditions in $x$- and $y$-direction, respectively (i.e. $T_{x/y} \Psi(\theta_x,\theta_y)=e^{i\theta_{x/y}}\Psi(\theta_x,\theta_y)$, where $T_{x/y}$ is a translation by the system size $L_{x/y}$ in $x$-, and $y$-direction, respectively). 

The twisted boundary conditions can be implemented in the same way as the magnetic flux discussed in Sec.\ \ref{sec:nonint} by transforming the hopping as $t_x\rightarrow t_x e^{i\theta_x/L_x}$ and $t_y\rightarrow t_y e^{i\theta_y/L_y}$ for hopping processes in $+\hat{x}$, and  $+\hat{y}$-direction, respectively, while taking the complex-conjugate in the opposite directions. The interaction and chemical potential terms in Eq.\ (\ref{eq:HofHarp}) remain unchanged. The only effect of adding the twisting angles $(\theta_x,\theta_y)$ to the infinite system is, as in Sec.\ \ref{sec:nonint},
\begin{equation}
\vec{v}_{k,q}\rightarrow \vec{v}_{k-\theta_x/L_x,q-\theta_y/L_y}.\nonumber
\end{equation}

In other words, if $T_{\theta_x,\theta_y}$ is the momentum-space translation operator which transforms each momentum as $k\rightarrow k+\theta_x/L_x$, and $q\rightarrow q+\theta_y/L_y$, we have
\begin{equation}
\left|\Psi(\theta_x,\theta_y)\right\rangle=T_{\theta_x,\theta_y}\left|\Psi(0,0)\right\rangle.\nonumber
\end{equation}
For the Berry-curvature
\begin{align}
\mathcal{B}\left(\theta_x,\theta_y\right) =& \partial_{\theta_x}\mathcal{A}_{y}-\partial_{\theta_y}\mathcal{A}_{x}  \nonumber\\
=& i\left(\left\langle\partial_{\theta_x}\Psi(\theta_x,\theta_y)\right|\partial_{\theta_y}\Psi(\theta_x,\theta_y)\rangle\right. \nonumber \\
& -\left.\left\langle\partial_{\theta_y}\Psi(\theta_x,\theta_y)\right|\partial_{\theta_x}\Psi(\theta_x,\theta_y)\rangle\right), \nonumber
\end{align}
we therefore have
\begin{align}
&\left\langle\partial_{\theta_i}\Psi(\theta_x,\theta_y)\right|\partial_{\theta_j}\Psi(\theta_x,\theta_y)\rangle= \nonumber\\
&\left[\left\langle\Psi(0,0)\right|\partial_{\theta_i}T^{\dagger}_{\theta_x,\theta_y}\right]\left[\partial_{\theta_j}T^{}_{\theta_x,\theta_y}\left|\Psi(0,0)\right\rangle\right]. \nonumber
\end{align}
The Berry curvature is therefore fully determined by the response of the periodic-boundary many-body groundstate $\Psi(0,0)$ to a translation in momentum.

If we define $P_{\rm h.c.}$ as the projector onto the Hilbert space of hard-core bosons (where multiple occupancy in position space is forbidden), the interacting many-body Hamiltonian [Eq.\ (\ref{eq:HofHarp})] can be written as
\begin{align}
H &= P_{\rm h.c.}\left(H_{\Phi}-\mu N\right)P_{\rm h.c.} \nonumber\\
&= \int dk dq \vec{v}_{k,q}\cdot P_{\rm h.c.}\vec{h}_{k,q}P_{\rm h.c.}-\mu P_{\rm h.c.}N P_{\rm h.c.}^{}, \nonumber
\end{align}
with particle-number operator $N$. The full momentum dependence of the hard-core bosons is therefore contained in the term $P_{\rm h.c.}\vec{h}_{k,q}P_{\rm h.c.}^{}$. Furthermore, for any hard-core boson many-body eigenstate $\Psi$ we have $P_{\rm h.c.}^{}\left|\Psi\right\rangle=\left|\Psi\right\rangle$. As in the non-interacting case therefore a non-trivial winding of $\left\langle \Psi(0,0)\left| \vec{h}_{k,q} \right|\Psi(0,0)\right\rangle$ in  momentum space indicates a non-trivial topology of the many-body groundstate. It should be emphasized that this measure is different from summing over the individual single-particle Chern numbers of the occupied bands, since no projection onto non-interacting bands is involved. 

For a further discussion of the measurement of topological properties with the $\hat{h}_{k,q}$-vector, see Appendix \ref{Chern}.

\section{Reciprocal Cluster Mean Field}
\label{Sec_RCMF}

The previously employed CGMF method \cite{Clust_MF,Natu_HHM} breaks translational invariance by applying the mean-field decoupling approximation only to the hopping-terms at the boundary of the cluster, while the hopping terms within the cluster are treated exactly. The simplest case where this can be observed is when the symmetry-breaking field is zero, reducing the lattice to a set of decoupled clusters with open boundaries. This violation of translational invariance breaks the symmetries of the dispersion and thereby its topological properties. In order to mitigate such artifacts we develop a mean-field decoupling based on the concept of momentum coarse-graining, introduced in the context of the dynamical cluster approximation \cite{Quant_Clust}.

We term this method as ``reciprocal cluster mean field" (RCMF). It crucially preserves both the translational invariance and the topology of the system. For topologically trivial translationally-invariant systems it yields more accurate results than previous mean-field methods (see Appendix \ref{DCAMF}).  It is well-suited for cases where the underlying symmetries of the dispersion are indispensable to understand the physical properties, such as, e.g., topological insulators. For benchmarks of the method on the hard-core Bose-Hubbard model and the chiral ladder with artificial magnetic fields \cite{Lad_Belen,Lad_Marie}, see Appendix \ref{DCAMF}.

To illustrate our procedure let us first start from a general non-interacting hopping Hamiltonian
\begin{equation}
H_{0}=\sum_{x',y'}\sum_{x,y}t_{\text{\tiny $(x',y')$,$(x,y)$}}b^{\dagger}_{\text{\tiny x',y'}}b_{\text{\tiny x,y}}=\sum_{k,q}\epsilon_{\text{\tiny k,q}}b^{\dagger}_{\text{\tiny k,q}}b_{\text{\tiny k,q}}
\label{hop_ham}
\end{equation}
with hopping amplitudes $t_{\text{\tiny $(x',y')$,$(x,y)$}}$ in position-space and dispersion $\epsilon_{\text{\tiny k,q}}$ in reciprocal space.
As in the dynamical cluster approximation \cite{Quant_Clust}, the main idea of RCMF consists in projecting the $N\times M$ lattice system onto a lattice of $N_c\times M_c$ clusters  (later we will take $N,M\rightarrow\infty$, but the method is also well-defined for finite systems). Each cluster is spanned by the internal cluster coordinates $X$ and $Y$, such that we can decompose the position coordinates $x$ and $y$ on the lattice into
\begin{equation}
x=X+\tilde{x}, \text{   }y=Y+\tilde{y}, \nonumber
\end{equation}
where $\tilde{x}$ and $\tilde{y}$ are inter-cluster coordinates. In the same way the momenta in $x$ and $y$-direction -- $k$ and $q$, respectively -- are decomposed as
\begin{equation}
k=K+\tilde{k}, \text{   } q=Q+\tilde{q},\nonumber
\end{equation}
where $K$ and $Q$ are the cluster momenta in reciprocal space. Through a partial Fourier transform, the creation and anihilation operators in reciprocal space can be written in the mixed representation
\begin{equation}
b^{}_{\text{\tiny $K+\tilde{k}$,$Q+\tilde{q}$}}=\frac{\sqrt{N_c M_c}}{\sqrt{NM}}\sum_{\tilde{x},\tilde{y}}e^{-i\left(\tilde{k}\tilde{x}+\tilde{q}\tilde{y}\right)}b_{\text{\tiny K,Q}}(\tilde{x},\tilde{y}),
\label{eq:op}
\end{equation}
where $b_{\text{\tiny K,Q}}(\tilde{x},\tilde{y})$ annihilates a boson with cluster-momenta $K$ and $Q$ on the cluster located at $(\tilde{x},\tilde{y})$ \cite{Quant_Clust}. 

The central idea of the momentum coarse-graining consists of projecting the dispersion of the lattice $\epsilon_{\text{\tiny k,q}}$ onto the clusters in reciprocal space. This can be done by a partial Fourier transform of the dispersion onto the subspace of cluster-local hopping processes, giving the intra-cluster dispersion $\bar{\epsilon}_{\text{\tiny K,Q}}$ as
\begin{equation}
\bar{\epsilon}_{\text{\tiny K,Q}}=\frac{N_cM_c}{NM}\sum_{\tilde{k},\tilde{q}}\epsilon_{\text{\tiny $K+\tilde{k}$,$Q+\tilde{q}$}},
\label{eq:epsilon_bar}
\end{equation}
representing hopping processes within the cluster, while the remainder $\delta \epsilon_{\text{\tiny K, $\tilde{k}$,Q, $\tilde{q}$}}=\epsilon_{\text{\tiny $K+\tilde{k}$,$Q+\tilde{q}$}}-\bar{\epsilon}_{\text{\tiny K,Q}}$  represents all other hopping processes between different clusters \cite{Quant_Clust}.

Now we can decompose the Hamiltonian of Eq.\ (\ref{hop_ham}) into
\begin{equation}
H_{0}=H_{\rm c}+\Delta H,
\label{eq:Sep_Ham}
\end{equation}
where, using Eq.\ (\ref{eq:op}), the part $H_{\rm c}$ is cluster-local,
\begin{align}
H_{\rm c} &=   \sum_{\tilde{k},\tilde{q}}\sum_{K,Q}\bar{\epsilon}_{\text{\tiny K,Q}}b^{\dagger}_{\text{\tiny K+$\tilde{k}$,Q+$\tilde{q}$}}b_{\text{\tiny K+$\tilde{k}$,Q+$\tilde{q}$}}\nonumber\\
& = \sum_{\tilde{x},\tilde{y}}\sum_{K,Q}\bar{\epsilon}_{\text{\tiny K,Q}}b^{\dagger}_{\text{\tiny K,Q}}(\tilde{x},\tilde{y})b_{\text{\tiny K,Q}}(\tilde{x},\tilde{y}),\nonumber
\end{align}
while $\Delta H$ contains the coupling between different clusters
\begin{align}
&\Delta H = \sum_{\tilde{k},\tilde{q}}\sum_{K,Q}\delta \epsilon_{\text{\tiny K, $\tilde{k}$,Q, $\tilde{q}$}}b^{\dagger}_{\text{\tiny $K+\tilde{k}$,$Q+\tilde{q}$}}b_{\text{\tiny $K+\tilde{k}$,$Q+\tilde{q}$}} \label{eq:DeltaH}\\
 &= \sum_{K,Q}\sum_{\tilde{x},\tilde{y}}\sum_{\tilde{x}',\tilde{y}'}\delta {\epsilon}_{\text{\tiny K,Q}}(\tilde{x}-\tilde{x}',\tilde{y}-\tilde{y}')b^{\dagger}_{\text{\tiny K,Q}}(\tilde{x},\tilde{y})b_{\text{\tiny K,Q}}(\tilde{x}',\tilde{y}'),\nonumber
\end{align}
where in the second line we introduced the mixed representation of $\delta \epsilon_{\text{\tiny K, $\tilde{k}$,Q, $\tilde{q}$}}$,
\begin{equation}
\delta {\epsilon}_{\text{\tiny K,Q}}(\tilde{x},\tilde{y})=\sum_{\tilde{k},\tilde{q}}e^{i\left(\tilde{k}\tilde{x}+\tilde{q}\tilde{y}\right)}\delta \epsilon_{\text{\tiny K, $\tilde{k}$,Q, $\tilde{q}$}}.\nonumber
\end{equation}

Our goal is to derive an effective Hamiltonian which is cluster local through a mean-field decoupling approximation of $\Delta H$. To this end we decompose the creation/annihilation operators into their static expectation values and fluctuations, i.e.
\begin{equation}
b_{\text{\tiny K,Q}}(\tilde{x},\tilde{y})=\phi_{\text{\tiny K,Q}}(\tilde{x},\tilde{y})+\delta b_{\text{\tiny K,Q}}(\tilde{x},\tilde{y}),
\label{decoup}
\end{equation}
where $\phi_{\text{\tiny K,Q}}(\tilde{x},\tilde{y})=\langle b_{\text{\tiny K,Q}}(\tilde{x},\tilde{y})\rangle$.

The standard procedure of the mean-field decoupling approximation consists of neglecting quadratic fluctuations.
Furthermore, we assume translational invariance between the different clusters, i.e. that the condensate $\phi_{\text{\tiny K,Q}}$ is independent of the cluster location
\begin{equation}
\phi_{\text{\tiny K,Q}}(\tilde{x},\tilde{y})=\phi_{\text{\tiny K,Q}}.\nonumber
\end{equation}
As derived in detail in Appendix \ref{Decoup}, this approach reduces a general system with local interactions and Hamiltonian
\begin{equation}
H^{'}=H_{0}+\frac{U}{2}\sum_{x,y} n_{\text{\tiny $x,y$}}\left(n_{\text{\tiny $x,y$}}-1\right)-\mu\sum_{x,y}n_{\text{\tiny $x,y$}},\nonumber
\label{eq:H_int}
\end{equation}
into a set of $\left(NM\right)/\left(N_c M_c\right)$ identical $N_c\times M_c$ cluster local systems with effective mean-field Hamiltonian,
\begin{align}
H^{'}_{\rm eff}= &\sum_{X',Y'}\sum_{X,Y}\bar{t}_{\text{\tiny $(X',Y')$,$(X,Y)$}}b^{\dagger}_{\text{\tiny $X'$, $Y'$}}b^{}_{\text{\tiny $X$, $Y$}}\nonumber \\
&-\mu\sum_{X,Y}n_{\text{\tiny $X,Y$}}+\frac{U}{2}\sum_{X,Y} n_{\text{\tiny $X,Y$}}\left(n_{\text{\tiny $X,Y$}}-1\right)\nonumber\\
&+\sum_{X,Y}\left(b^{\dagger}_{\text{\tiny $X$, $Y$}}F_{\text{\tiny $X$, $Y$}}+F^{*}_{\text{\tiny $X$, $Y$}}b^{}_{\text{\tiny $X$, $Y$}}\right).
\label{eq:H11_mf}
\end{align}
The symmetry breaking field $F_{\text{\tiny $X$, $Y$}}$ is given by
\begin{equation}
F_{\text{\tiny $X$, $Y$}}=\sum_{X',Y'}\delta t_{\text{\tiny $(X,Y)$,$(X',Y')$}}\phi^{}_{\text{\tiny $X'$, $Y'$}},\label{eq:F_xy}
\end{equation}
and the effective hopping amplitudes are defined as
\begin{align}
& \bar{t}_{\text{\tiny $(X',Y')$,$(X,Y)$}}=\frac{1}{N_c M_c}\sum_{K,Q}e^{\text{\tiny $i\left(K\left(X'-X\right)+Q\left(Y'-Y\right)\right)$}}\bar{\epsilon}_{\text{\tiny $K,Q$}}, \nonumber\\
& \delta t_{\text{\tiny $(X',Y')$,$(X,Y)$}}=t_{\text{\tiny $(X',Y')$,$(X,Y)$}}-\bar{t}_{\text{\tiny $(X',Y')$,$(X,Y)$}}. \label{eq:delta_t}
\end{align}
Within the RCMF approach the effective free energy of the lattice system is given by
\begin{equation}
\Omega = \Omega^{'}-\frac{1}{2}\sum_{X,Y}\left(\phi^{*}_{\text{\tiny $X$, $Y$}} F_{\text{\tiny $X$, $Y$}}+F^{*}_{\text{\tiny $X$, $Y$}}\phi^{}_{\text{\tiny $X$, $Y$}}\right),
\label{eq:Omega1}
\end{equation}
where $\Omega^{'}$ is the free energy of the cluster local Hamiltonian of Eq.\ (\ref{eq:H11_mf}). Note that Eq.\ (\ref{eq:Omega1}) is consistent with the standard lattice free-energy within the single-site mean-field approximation \cite{BSFT}. 
In fact, requiring stationarity in the symmetry breaking field  $F_{\text{\tiny $X$, $Y$}}$, 
\begin{equation}
\frac{\delta\Omega}{\delta F_{\text{\tiny $X$, $Y$}}}=\frac{\delta\Omega}{\delta F^{*}_{\text{\tiny $X$, $Y$}}}=0,\nonumber
\label{eq:Stat}
\end{equation}
taking into account Eq.\ (\ref{eq:F_xy}), reproduces the standard mean-field self-consistency condition
\begin{equation}
\phi^{}_{\text{\tiny $X$, $Y$}}=\langle b_{\text{\tiny $X$, $Y$}}\rangle.
\label{eq:SelfCon}
\end{equation}
Here, $\langle.\rangle$ means taking the expectation value with respect to the mean-field Hamiltonian [Eq.\ (\ref{eq:H11_mf})].

We note in passing that the treatment of the symmetry-breaking field $F$ is identical to the way it should be implemented in a dynamical cluster approximation extension of bosonic dynamical mean-field theory \cite{Byc_BDMFT,BDMFT1,BDMFT}.

\section{Results}
\label{Sec_Res}
\begin{figure}
  \includegraphics[scale=1]{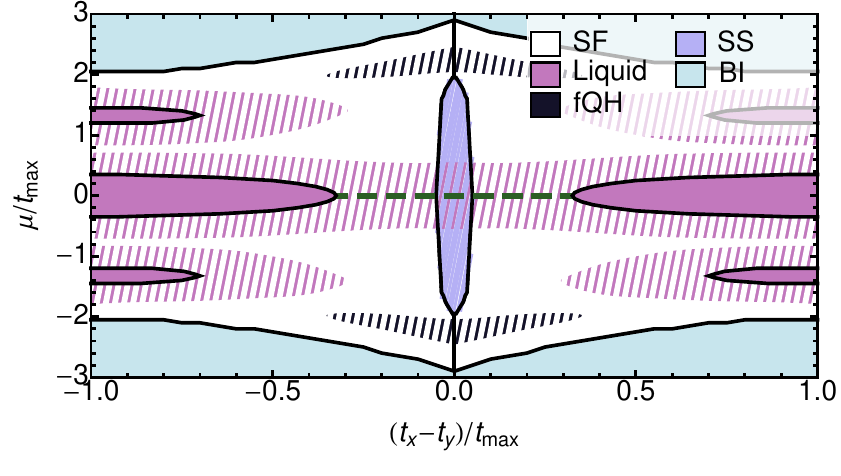}
  \caption{\label{fig:PhaseDiagram} Groundstate phase diagram of the HHMm in two dimensions with hard-core bosons and flux $\Phi=\pi/2$ in terms of $\mu/t_{\rm max}$ and $\left(t_x-t_y\right)/t_{\rm max}$.
    The observed phases are band insulating (BI, light blue), supersolid (SS, dark blue), striped superfluid (SF, white), gapless uncondensed liquid (Liquid, pink), and fractional quantum Hall (fQH, dark gray). The dashed regions indicate where the RCMF groundstate has a non-zero condensate order parameter but is very close in energy ($<3\%$) to metastable uncondensed states.
At zero anisotropy the striped superfluid undergoes phase separation between vertically (for $t_x>t_y$) and horizontally (for $t_x<t_y$) striped order (black vertical line), while for $\mu=0$ the density is homogeneous and fixed to $n=1/2$ in all phases (green dashed line).}
\end{figure}

In our RCMF approach the HHMm [Eq.\ (\ref{eq:HofHarp})] is reduced to an effective $4\times 4$ cluster Hamiltonian, for details see Appendix \ref{hhmm}. For a comparison of our RCMF results with exact diagonalization at zero hopping anisotropy, see Appendix \ref{Comp_ED}. In Fig.\ \ref{fig:PhaseDiagram} we present the groundstate phase diagram in terms of the chemical potential $\mu/t_{\rm max}$ and the hopping-anisotropy $(t_x-t_y)/t_{\rm max}$, where $t_{\rm max}={\rm max}\left[\lbrace t_x, t_y\rbrace\right]$. The phases at densities $n$ and $1-n$  are related by a $\mathcal{C}\mathcal{T}$-transformation consisting of a particle-hole transform combined with complex-conjugation (see Appendix \ref{cchb}). The symmetry around the $(t_x-t_y)=0$ axis corresponds to gauge invariance, since $t_x$ and $t_y$ can be exchanged in combination with a lattice-rotation of $\pi/2$. At $n=0$ and $n=1$ we find topologically trivial band insulators (BI). Below we discuss the other resulting phases in more detail.

\subsection{Condensed phases}

\begin{figure}
  \includegraphics[scale=.85]{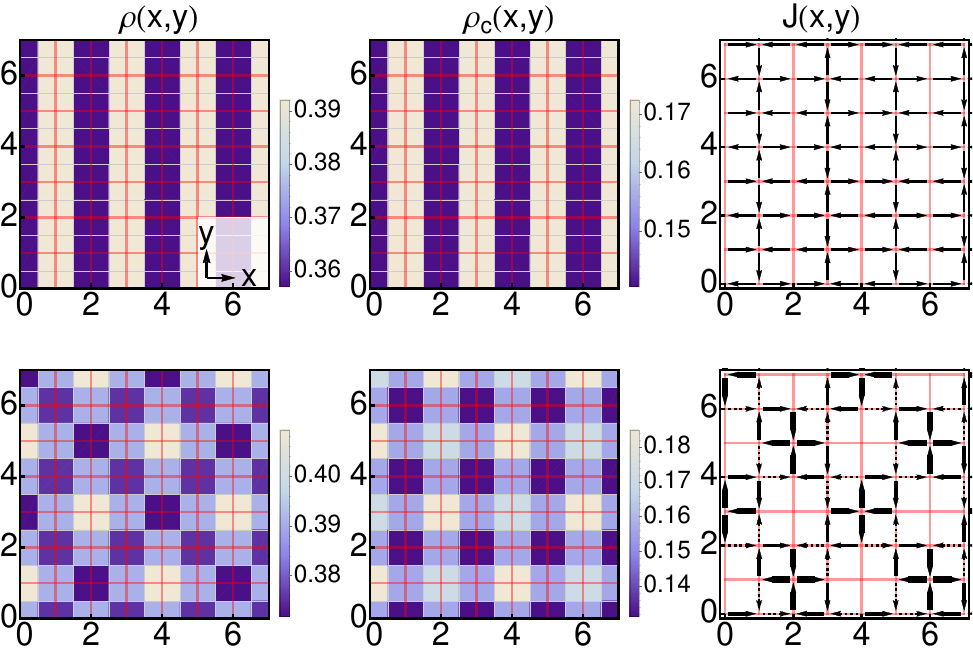}
  \caption{\label{fig:Points} Density (left column), condensate-density (central column), and current patterns (right column) for the VS-SF at $\mu/t_{\rm max}=-0.8$, $(t_x-t_y)/t_{\rm max}=0.2$ (upper row), and the SS at $\mu/t_{\rm max}=-0.8$, $(t_x-t_y)/t_{\rm max}=0$ (lower row).
The arrow-thickness indicates the magnitude of the currents. }    
\end{figure}
At moderate values of $\mu$ we observe superfluid phases with striped density and condensate density modulation. For $t_x>t_y$ this is a vertically striped superfluid (VS-SF), with vertically striped density distribution $\rho(x,y)$ and condensate-density distribution $\rho_c(x,y)=|\phi_{x,y}|^2$, as shown in Fig.\ \ref{fig:Points} together with the spatially resolved particle current $\vec{J}(x,y)$. The net current is zero, as expected for an infinite system. Locally, however, there are chiral currents around two plaquettes in the horizontal direction. We therefore introduce the striped-superfluid order parameter $J_{\rm str}=\sum_{x,y}\left[\cos\left(\frac{\pi}{2}(x+2y)\right)  J_{\rm x}(x,y)  -\cos\left(\frac{\pi}{2}(2x+y)\right)  J_{\rm y}(x,y) \right]$, where $J_{\rm x(y)}(x,y)$ is the groundstate expectation value of the  current in $x$ ($y$) direction.
For $t_x<t_y$ the superfluid phase is horizontally striped (HS-SF), with the patterns of Fig.\ \ref{fig:Points} rotated by $\pi/2$ compared to the VS-SF. Since at $t_x=t_y$ the system is invariant under a $\pi/2$-rotation, for $|\mu|/t_{\rm max}\gtrsim 2$ the VS-SF and HS-SF undergo phase separation. 

At $|\mu|/t_{\rm max}\lesssim 2$ and for low anisotropy we find a supersolid phase (SS) with lower free energy than the striped phases. The density distributions $\rho$ and $\rho_c$ spontaneously break translational invariance, having a  period larger than the unit cell (see Fig.\ \ref{fig:Points}). A similar spontaneous breaking of translational invariance has already been observed in the staggered-flux bosonic Harper-Mott model \cite{Mol_SS} and the bosonic Hofstadter model on a dice lattice \cite{Mol_SS_Dice}, and has recently been measured experimentally in spin-orbit coupled Bose-Einstein condensates \cite{Kett_SS}. The SS exhibits chiral currents around single plaquettes, with position-dependent amplitudes, as captured by the order parameter $J_{\rm ss}=\sum_{x,y}\cos\left(\frac{\pi}{2}(x+y)\right)\left(J_{\rm x}(x,y)-J_{\rm y}(x,y)\right)$.
In all phases, at $\mu=0$ the density distribution is homogeneous, $\rho(x,y)=1/2$, while $\rho_c(x,y)$ remains modulated.
\begin{figure}
  \includegraphics[scale=.85]{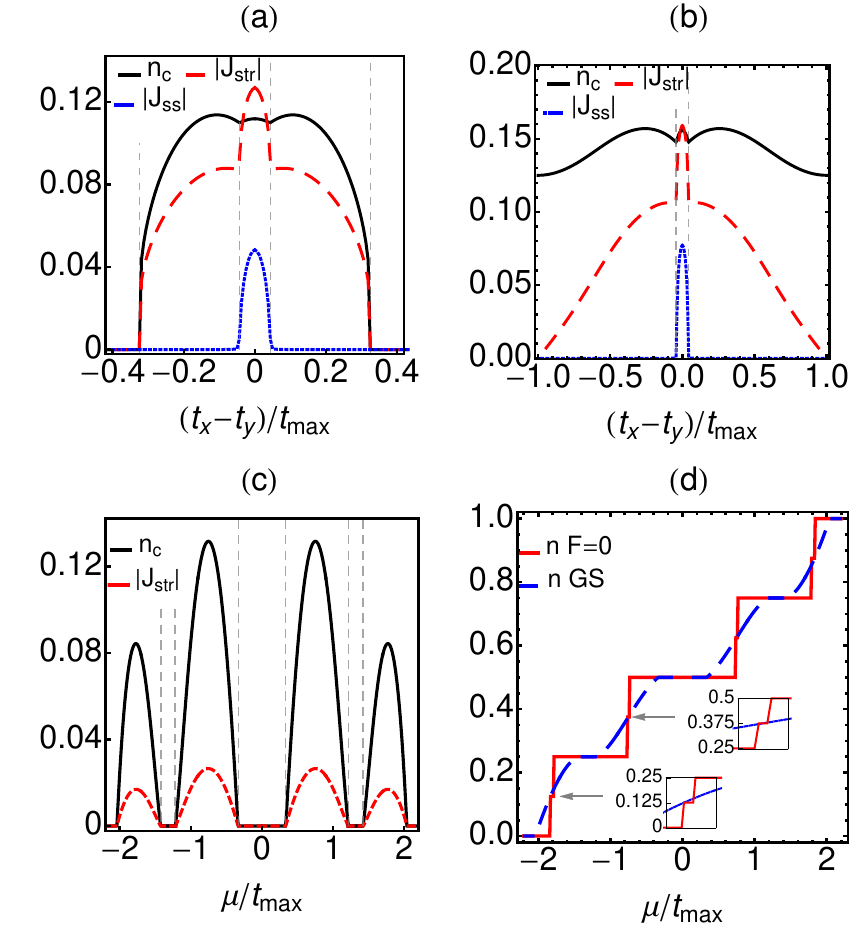}
  \caption{\label{fig:Ani_0} Order parameters and densities. (a) and (b) Sweep in anisotropy at fixed $\mu/t_{\rm max}$. In (a) the average condensate density $n_c$ (black), and the order parameters $J_{\rm str}$ (red, dashed), and $J_{\rm ss}$ (blue, dotted) are shown for $\mu/t_{\rm max}=0$. In (b) the same quantities are shown for $\mu/t_{\rm max}=-0.8$. (c) and (d) Sweep in $\mu$ at fixed $(t_x-t_y)/t_{\rm max}=-0.8$. In (c) $n_c$ (black) and $J_{\rm str}$ (red, dashed) are shown. In (d) the average density $n$ is shown in the groundstate (blue dashed) and for the stationary solution with zero symmetry-breaking field $F$ (red). The insets indicate the regions where the $F=0$ solution shows plateaus at fractional filling $\nu=1/2$ ($n=1/8$) and $\nu=3/2$ ($n=3/8$), respectively. In (a), (b), and (c) the vertical dashed lines indicate phase transitions.
  }    
\end{figure}

The phase transition between the striped superfluids and the SS phase is characterized by a kink in the average condensate density $n_c$, see Figures \ref{fig:Ani_0}a and  \ref{fig:Ani_0}b. For $n_c>0$, the striped superfluid order parameter $J_{\rm str}$ is only zero at $|t_x-t_y|/t_{\rm max}=1$ (where the lattice is a set of trivial one-dimensional chains), exhibiting a kink at the phase transition to the SS, where also $J_{\rm ss}$ becomes non-zero.

\subsection{Uncondensed phases}
\label{incom}

At density $n=1/2$ (Fig.\ \ref{fig:Ani_0}a) and stronger anisotropy we find a phase with zero condensate density ($n_c=0$). In Figs.\ \ref{fig:Ani_0}c and \ref{fig:Ani_0}d we show a sweep in $\mu$ for $(t_x-t_y)/t_{\rm max}=-0.8$, where we observe plateaus with zero $n_c$, zero current, and homogeneous density distribtution $\rho(x,y)=\nu/4$, with fillings $\nu=1,2,3$. In these phases, since $F_{x,y}=0$, the RCMF Hamiltonian of Eq.\ (\ref{eq:H11_mf}) reduces to a finite $4\times 4$ torus without any external variational parameter. In order to further analyze these phases we therefore turn to ED using twisted boundary conditions in order to analyze finite-size effects (see Sec.\ \ref{sec:interact} and Appendix \ref{Comp_ED}). If the phases are gapped, one expects the manybody gap to stay finite for all twisting angles $\left(\theta_x,\theta_y\right)$, while in gapless phases the groundstate mixes with excited states.
\begin{figure}
  \includegraphics[scale=.48]{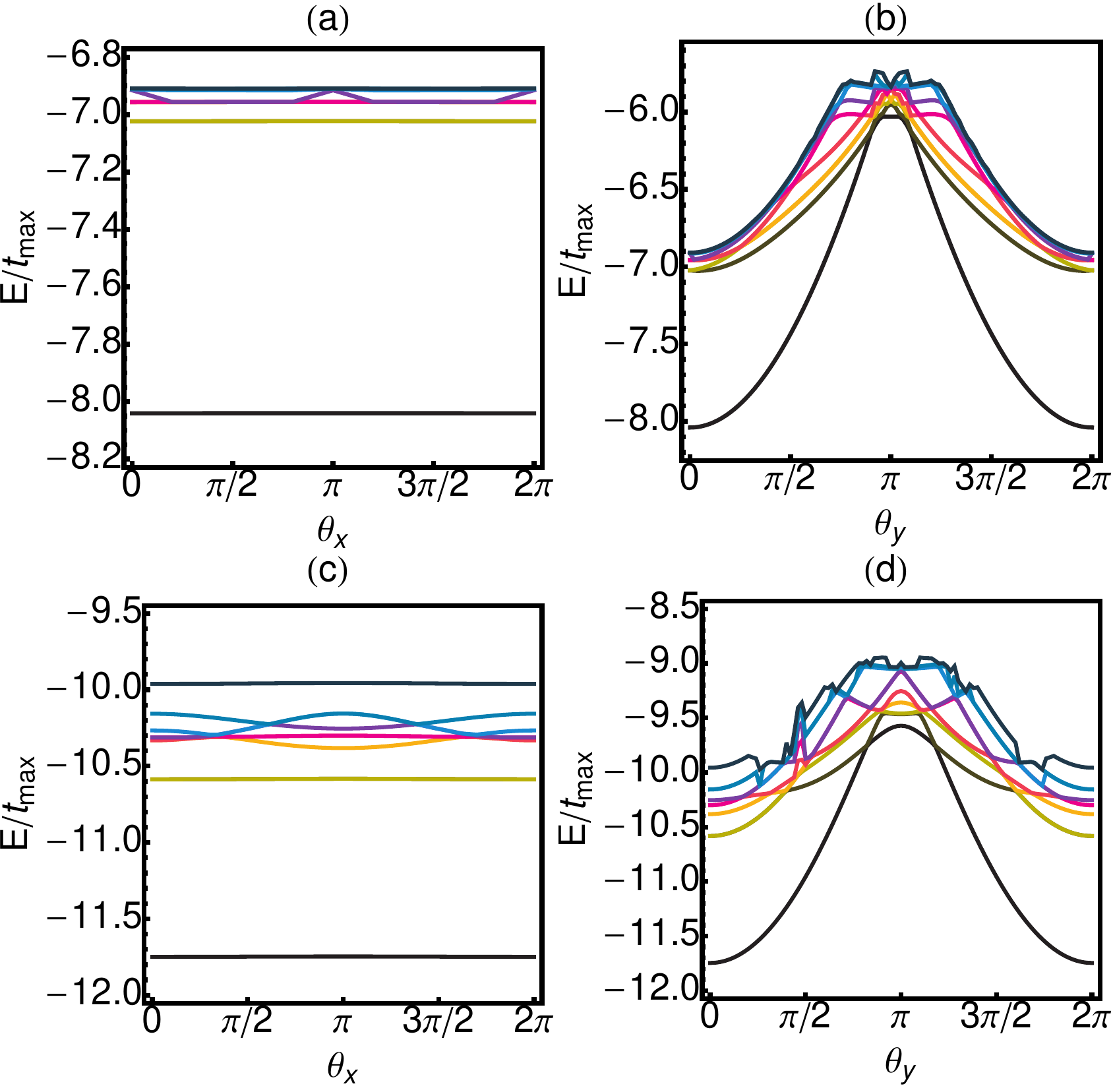}
  \caption{\label{fig:Twist} Response of the lowest 10 eigenvalues of a $4\times 4$ system to twisted boundary conditions. (a) and (b): For $4$ particles ($\nu=1$), $t_x=0.2$, $t_y=1$, and fixed twisting angles $\theta_y=0$ (a) and $\theta_x=0$ (b).  (c) and (d): For $8$ particles ($\nu=2$), $t_x=0.5$, $t_y=1$, and fixed twisting angles $\theta_y=0$ (c) and $\theta_x=\pi$ (d).}
\end{figure}

As can be seen in Fig.\ \ref{fig:Twist} for $t_x<t_y$ the groundstate remains gapped with respect to boundary twisting in the $x$-direction with $\theta_y=0$, while it mixes with the excited states for twisting in the $y$-direction. For $t_y<t_x$ the behavior is reversed. This is also consistent with the correlations $|\langle b^{\dagger}_{x,y} b_{0,y}\rangle|$ decreasing exponentially to zero as a function of $x$ for $t_x<t_y$, while staying finite throughout the system for $t_x>t_y$ (see Appendix \ref{Comp_ED}). In contrast to the two-dimensional Bose-Hubbard model without magnetic flux, which in the superfluid groundstate always shows condensation as long as both hopping amplitudes are finite \cite{QMC_2D_ani}, the HHMm therefore shows a transition at finite $t_x/t_y$ to a highly anisotropic uncondensed gapless liquid. The fact that these phases are adiabatically connected to the one-dimensional limit ($t_x=0$ or $t_y=0$), where hard-core bosons are in a superfluid phase, as well as the highly anisotropic correlations $|\langle b^{\dagger}_{x,y} b_{0,y}\rangle|$, possibly point to an unconventional one-dimensional superfluid order.
\begin{figure}
  \includegraphics[scale=.93]{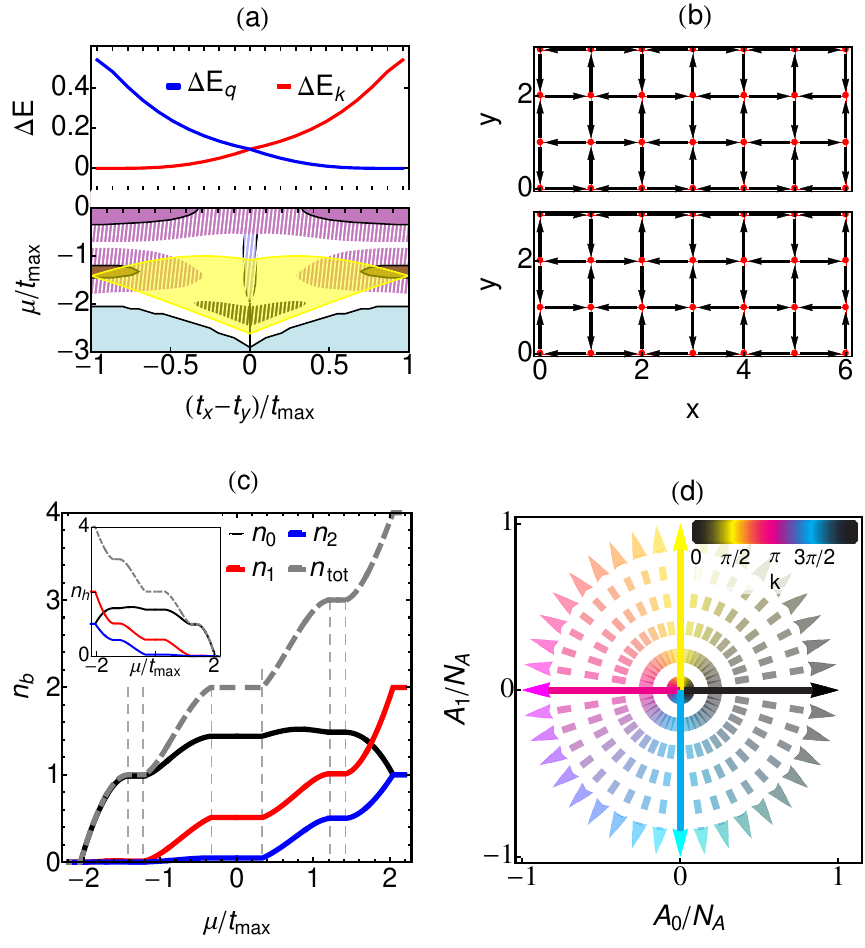}
  \caption{\label{fig:iQHE} Uncondensed phases. (a) top: Bandwidths of the lowest band in $k$-, and $q$-direction, $\Delta E_k$ (red) and  $\Delta E_q$ (blue), as a function of $(t_x-t_y)/t_{\rm max}$; (a) bottom: quantum Hall plateau for non-interacting fermions (yellow) compared to the hard-core boson phase diagram. (b) Two counter-propagating current patterns (upper and lower panel, respectively) whose sum gives zero net current, resulting from current-current correlations.  (c) Occupations of the lowest ($n_0$, black), central ($n_1$, red), highest ($n_{2}$, blue) band, and total occupation $n_{\rm tot}=n_0+n_1+n_2$ (gray dashed), for $(t_x-t_y)/t_{\rm max}=-0.8$ as a function of $\mu$. The phase transitions between condensed and uncondensed phases are indicated with dashed vertical lines. In the inset the corresponding hole occupations ($n_{h}=\langle b b^{\dagger}\rangle$)  are shown in the same colors. (d) $A_0$ and $A_1$ components of the $\hat{h}_{k,q}$ vector [Eq.\ (\ref{eq:Svec})] for $(t_x-t_y)/t_{\rm max}=-0.8$ as a function of $k$ (see coloring) in the single-particle case (dashed arrows), and for hard-core bosons with $\mu/t_{\rm max}=0$ (full arrows). $A_0$ and $A_1$ are normalized by $N_A=\sqrt{A_0^2+A_1^2}$, while the $B$-component varies only slightly (not shown). }    
\end{figure}

As a function of the hopping anisotropy, the liquid phases occur where the lowest band is particularily flat either in $k$- or  $q$-direction, suppressing condensation in the minima of the dispersion (see Fig.\ \ref{fig:iQHE}a and Appendix \ref{hhm}).
While the system has zero current everywhere (due to the periodic boundaries), a signature of the response of the liquid to the magnetic field is found by analyzing current-current correlations (see Appendix \ref{Sec:cur_cur} for details), resulting in two counter-propagating currents which cancel each other, shown in Fig.\ \ref{fig:iQHE}b.

In Fig.\ \ref{fig:iQHE}c we show the projection of the groundstate onto the three non-interacting bands $n_0$, $n_1$, and $n_2$, for the same parameters as in Figs.\  \ref{fig:Ani_0}c and \ref{fig:Ani_0}d. At $\nu=1$ the lowest band shows unit filling. As shown in Fig.  \ref{fig:iQHE}a, this phase appears in the same regions of $\mu$ as the integer quantum Hall plateau of non-interacting spinless fermions (see Appendix \ref{hhm})). At $\nu=3$ the holes show unit filling in the lowest band, due to the $\mathcal{C} \mathcal{T}$ transform (see Appendix \ref{cchb}).
\begin{figure*}
  \includegraphics[scale=.425]{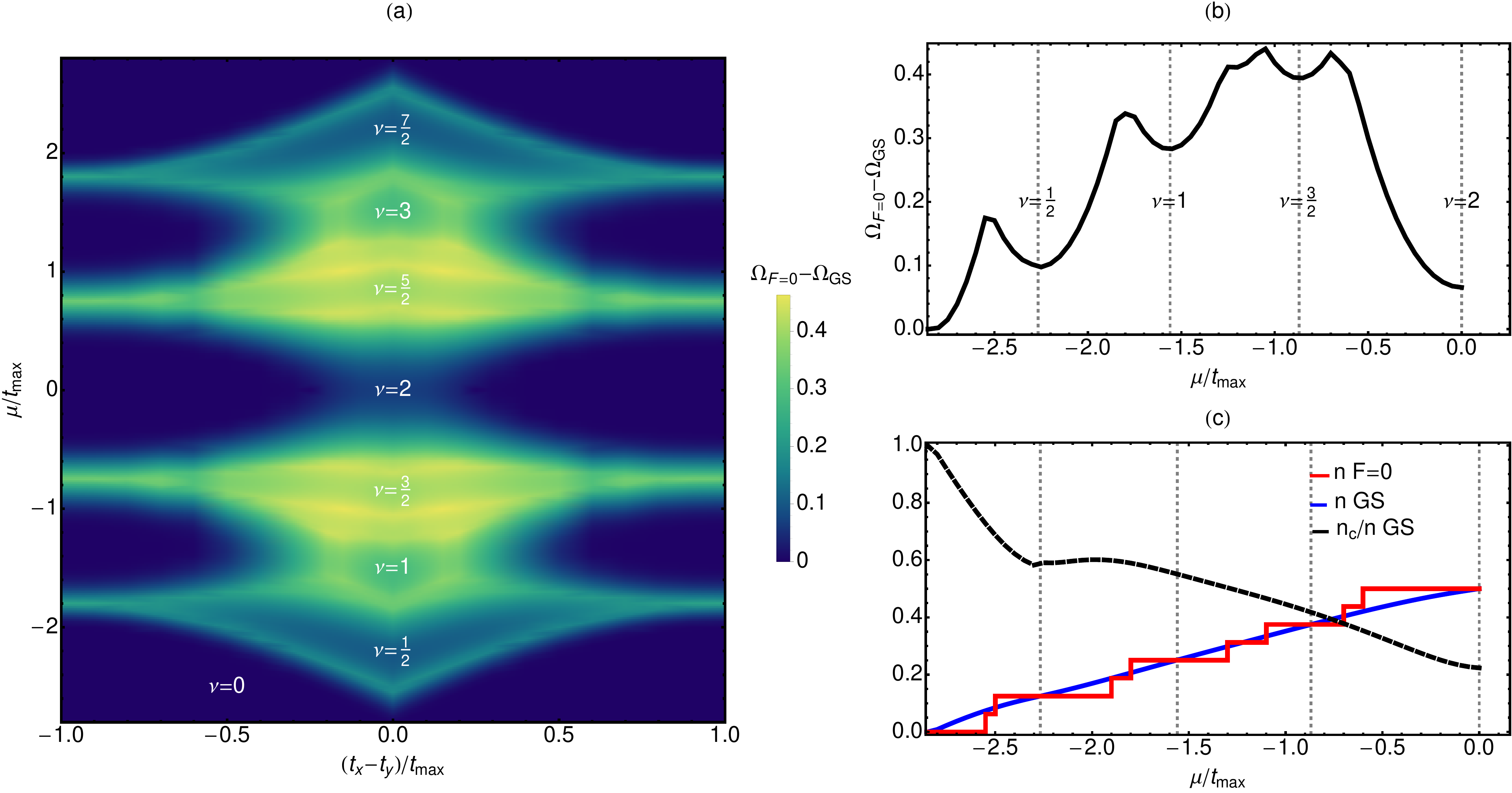}
  \caption{\label{fig:EnDif} Analysis of metastable phases. (a) and (b) Difference in free energy between the solution with zero symmetry-breaking field ($\Omega_{F=0}$) and the groundstate ($\Omega_{\rm GS}$) (a) as a function of anisotropy and chemical potential, (b) as a function of chemical potential at $t_x=t_y$. (c) Condensate fraction $n_c/n$ (blacked, dashed) and density (blue) of the groundstate compared to the density of the $F=0$ solution (red) as a function of chemical potential and $t_x=t_y$. In panels (b) and (c) the vertical dotted lines show the locations of the metastable plateaus, while in panel (a) they are indicated with white labels.}   
\end{figure*}

As can be seen in Fig.\ \ref{fig:iQHE}d, the vector $\hat{h}_{k,q}$ shows the same behavior as for the lowest non-interacting band in all three liquid phases (shown for $\nu=2$), in contrast to the trivial BI at $n=0,1$ and the one-dimensional superfluid at $t_x=0$ or $t_y=0$. For $\nu=1$, as in the case of non-interacting fermions discussed in Sec.\ \ref{sec:nonint}, this winding indicates the transverse transport of a single particle if a magnetic flux of $\Phi_{\rm y}=2\pi$ is inserted. At $\nu=3$, the transverse transport consists of a single hole. This is consistent with the band fillings in Fig.\ \ref{fig:iQHE}c and  the $\mathcal{C} \mathcal{T}$ transform discussed in Appendix \ref{cchb}, i.e. the reversal of the Hall conductivity $\sigma_{xy}(\nu=3)=-\sigma_{xy}(\nu=1)$. As these phases are gapless in the two-dimensional thermodynamic limit, the quantization of the Hall conductivity is not topologically protected by edge modes and therefore sensitive to disorder, as is the case for metallic Fermi-liquid-like phases of hard-core bosons \cite{Zeng_16,Read,Sheng_Dis}. 

By contrast, in the case of a cylindrical geometry, i.e. $L_y=4$ and $L_x\rightarrow\infty$, the response to the twisted boundaries in $x$-direction while $\theta_y=0$, shown in Figs.\ \ref{fig:Twist}a and  \ref{fig:Twist}c, indicates that all three plateaus are gapped. As a function of $\theta_x$ the vector $\hat{h}_{k,q}$ shows a complete loop and appears to be robust against local perturbations (see Appendix \ref{Chern}). What the nature of the phases at $\nu=1,3$ is in such a quasi-one-dimensional setup remains to be investigated: the non-trivial winding indicates a gapped phase with odd Hall conductivity, which is expected to show intrinsic topological order and fractional quasiparticle excitations for bosons in two dimensions \cite{Senthil,Lu_17}. The non-degenerate groundstate we observe (which for bosons is only expected at even Hall conductivities) apparently is at odds with this prediction. However, the argument of Ref.\ \onlinecite{Senthil} relies on the fact that the quasiparticle excitations need to behave as fermions under exchange in such an odd Hall conductivity phase. In the cylinder, however, we approach the one-dimensional limit, where hard-core bosons naturally behave as free fermions also in the absence of fractionalization. This possibly explains why we observe an ED groundstate which remains gapped and non-degenerate for all accessible system sizes $L_x\times 4$ (see Appendix \ref{Comp_ED}).

At $\nu=2$ ($n=1/2$), the Hamiltonian of Eq.\ (\ref{eq:HofHarp}) is $\mathcal{C} \mathcal{T}$-symmetric. It directly follows that $\sigma_{xy}(\nu=2)=0$. This is consistent with the bands being equally filled with particles and holes (see Fig.\ \ref{fig:iQHE}b), resulting in a zero net Hall conductivity (see Appendix \ref{cchb}). In two-dimensional systems such a $\mathcal{C} \mathcal{T}$-symmetric phase is expected to be topologically trivial \cite{Tenfold,XiaoGangWen_SPT_Science,Vish}, in line with the gaplessness observed in Figs.\ \ref{fig:Twist}c and \ref{fig:Twist}d.

On a cylinder, however, this phase is gapped as the one-dimensional limit is approached, where $\mathcal{C} \mathcal{T}$-symmetric phases can have a non-zero topological $\mathbb{Z}$ invariant for non-interacting fermions \cite{Tenfold}. The non-trivial winding implies the quantized transport of a single particle-hole pair under the adiabatic insertion of a flux of $2\pi$, resulting in a total zero Hall conductivity. Wether this particle-hole transport is a consequence of topologically protected edge modes on the cylinder remains to be investigated on larger system sizes, possibly using DMRG \cite{He_HHM,Pol_SPT,Motruk}.

Whereas away from integer fillings the groundstate is always symmetry-broken, it is always possible within RCMF to find (metastable) stationary solutions with zero symmetry-breaking field ($F=0$) and therefore $n_c=0$, as shown in Fig.\  \ref{fig:Ani_0}d. While at large hopping anisotropy fractional fillings are largely supressed, at low anisotropy the $F=0$ solution shows plateaus at any filling commensurate with the $4\times 4$ cluster, i.e. $\nu=m/4$ with integer $m$, as shown in Fig.\ \ref{fig:EnDif}c. 

As mean-field approaches such as RCMF tend to overestimate the stability of symmetry-broken phases, we critically analyze the difference in free energy between the groundstate and the metastable plateaus in Fig.\ \ref{fig:EnDif}. As can be seen, this quantity shows local minima at integer ($\nu=1,2,3$) and half-integer ($\nu=1/2,3/2,5/2,7/2$) fillings, while quarter fillings correspond to local maxima. This is consistent with the argument that without long-range interactions it costs a negligible energy to compress the $\nu=1/4$ to the $\nu=1/2$ Laughlin liquid \cite{ED_07}. The energy difference is particularily low in the vicinity of the liquid phases indicating that these plateaus might extend to lower values of hopping anisotropy. Furthermore, at low anisotropy (where the lowest band is particlularily flat, see Fig.\ \ref{fig:iQHE}c  and Appendix \ref{hhm}) the metastable plateau at $\nu=1/2$ (and $\nu=7/2$) is very close in free energy to the groundstate. This plateau has been shown to correspond to a fQH phase in ED \cite{ED_05,ED_07,Mol_CFM}, variational Gutzwiller mean field \cite{Onur2}, and DMRG \cite{He_HHM} studies. As shown in Fig.\ \ref{fig:EnDif}c, at zero anisotropy, the condensate fraction of the groundstate shows a local minimum at both $\nu=1/2$ and $\nu=2$, further indicating that it might converge to zero with increasing cluster size.
These are also the fillings where at zero anisotropy there is the largest discrepancy between RCMF and ED results on small finite systems  (see Appendix \ref{Comp_ED}). It should however be noted that when assuming a cylinder geometry, RCMF observes both a $\nu=1/2$ and $\nu=2$ plateau, in agreement with ED. 
To conclude, there are regions of the phase diagram, where the symmetry-broken groundstate and the metastable plateaus are too close in free energy to dismiss finite size effects. We denote these regions (identified by the condition $\left|\Omega_{\rm GS}-\Omega_{F=0}\right|<3\%$ of the groundstate energy) as dashed areas in the phase diagram of Fig.\  \ref{fig:PhaseDiagram}.

\section{Conclusion}
\label{Sec_Conc}

We derived the reciprocal cluster mean field method and applied it to the groundstate phase diagram of hard-core bosons in the Harper-Hofstadter-Mott model at flux $\Phi=\pi/2$.  The bosons exhibit band insulating, striped superfluid, and supersolid phases. At finite anisotropy and integer filling we further found anisotropic gapless uncondensed liquid groundstates characterized by a non-trivial winding of the newly introduced vector $\hat{h}_{k,q}$.  We further analyzed the properties of these phases using exact diagonalization. At fillings $\nu=1$ ($3$) this corresponds to integer particle (hole) filling of the lowest band, while the $\nu=2$ phase is $\mathcal{C}\mathcal{T}$ symmetric with zero Hall response. We also observed metastable fractional quantum Hall phases predicted by other methods \cite{ED_05,ED_07,Mol_CFM,Onur2,He_HHM}, which do not correspond to the groundstate (most likely due to finite-size effects), but are very close in free energy.
 Finally, we discussed how the liquid phases at integer fillings become gapped on a cylinder with just one unit-cell in the $y$-direction and show a quantized Hall response to the adiabatic insertion of a magnetic flux. These properties, which are not expected for the full two-dimensional system, seem inherent to the quasi-one-dimensional nature of the cylinder geometry and need to be further investigated on larger system sizes. The combination of reciprocal cluster mean-field and exact diagonalization provides a promising venue for the numerical simulation of bosonic lattice systems with larger unit cells in the thermodynamic limit.

\begin{acknowledgments}
The authors would like to thank I. Bloch, H. P. B\"uchler, N. R. Cooper, F. Grusdt, A. Hayward, Y.-C. He, F. Heidrich-Meisner, F. Kozarski, A. M. L\"auchli, K. Liu, M. Lohse, G. M\"oller, T. Pfeffer, M. Piraud, F. Pollmann, K. Sun, and R. O. Umucalilar, C.Wang, and N. Yao for valuable input. We especially thank A. M. L\"auchli and F. Pollmann for sharing unpublished data and critically analyzing the results at $\nu=1/2$ and Y.-C. He for pointing out symmetry constraints in the uncondensed phases. DH and LP are supported  by  FP7/ERC  Starting  Grant  No.   306897  and FP7/Marie-Curie Grant No.  321918, HS and PW by FP7/ERC starting grant No. 278023.
\end{acknowledgments}

\appendix

\section{Anisotropic Harper-Hofstadter model}
\label{hhm}

The HHm can be solved by diagonalizing the Hamiltonian of Eq.\ (\ref{eq:HofHarp_kq}), yielding three topologically non-trivial bands (see Fig 1b). For the gauge used in this work, the non-trivial topology arises in $k$-direction, while in $q$-direction the dispersion has a trivial cosine-shape. Both the topology and the four minima of the dispersion are independent of the anisotropy between the hopping amplitudes $t_x$ and $t_y$. The bandwidths of the three bands, on the other hand, are affected by the ratio $t_x/t_y$. 

In order to analyze this, we introduce the quantities $\Delta E_k$ and $\Delta E_q$ for the lowest band, where $\Delta E_k$ is the bandwidth in $k$-direction, i.e.
\begin{equation}
\Delta E_k = \max_{q}\Delta\tilde{E}_{k}(q),\nonumber 
\end{equation}
where
\begin{equation}
\Delta\tilde{E}_{k}(q) = \max_{k}\epsilon^{\text{\tiny 0}}(k,q)-\min_{k}\epsilon^{\text{\tiny 0}}(k,q), \nonumber
\end{equation}
$\epsilon^{\text{\tiny 0}}(k,q)$ is the dispersion of the lowest band,  and $\max_{k/q}$ corresponds to taking the maximum with respect to $k$ and $q$, respectively. The bandwidth in $q$-direction,  $\Delta E_q$ is defined analogously. As shown in the upper panel of Fig.\ \ref{fig:iQHE}c, for $t_x\ll t_y$ the lowest band is particularily flat in the $k$-direction ($\Delta E_k\ll 1$), while for $t_y\ll t_x$ it is particularily flat in the $q$-direction ($\Delta E_q\ll 1$). Note that this is not to be confused with the "flatness" of the bands that typically supports fractional quantum Hall effects, which would consist in $\rm{max}\left[\Delta E_k,\Delta E_q\right]\ll 1$ (in fact this quantity is low in the region of low anisotropy). Instead, having only $\Delta E_k\ll 1$ or $\Delta E_q\ll 1$ will result simply in supressing the condensation of bosons in the minima of the dispersion.

Another quantity affected by the anisotropy is the gap between the lowest and the central band. The simplest many-body problem where this plays a role is the case of spinless non-interacting fermions, which exhibit an integer quantum Hall phase for integer filling of the lowest band, i.e. if the chemical potential $\mu$ lies within the (anisotropy-dependent) gap, see Fig.\ 5c.
\begin{figure}
  \includegraphics[scale=.38]{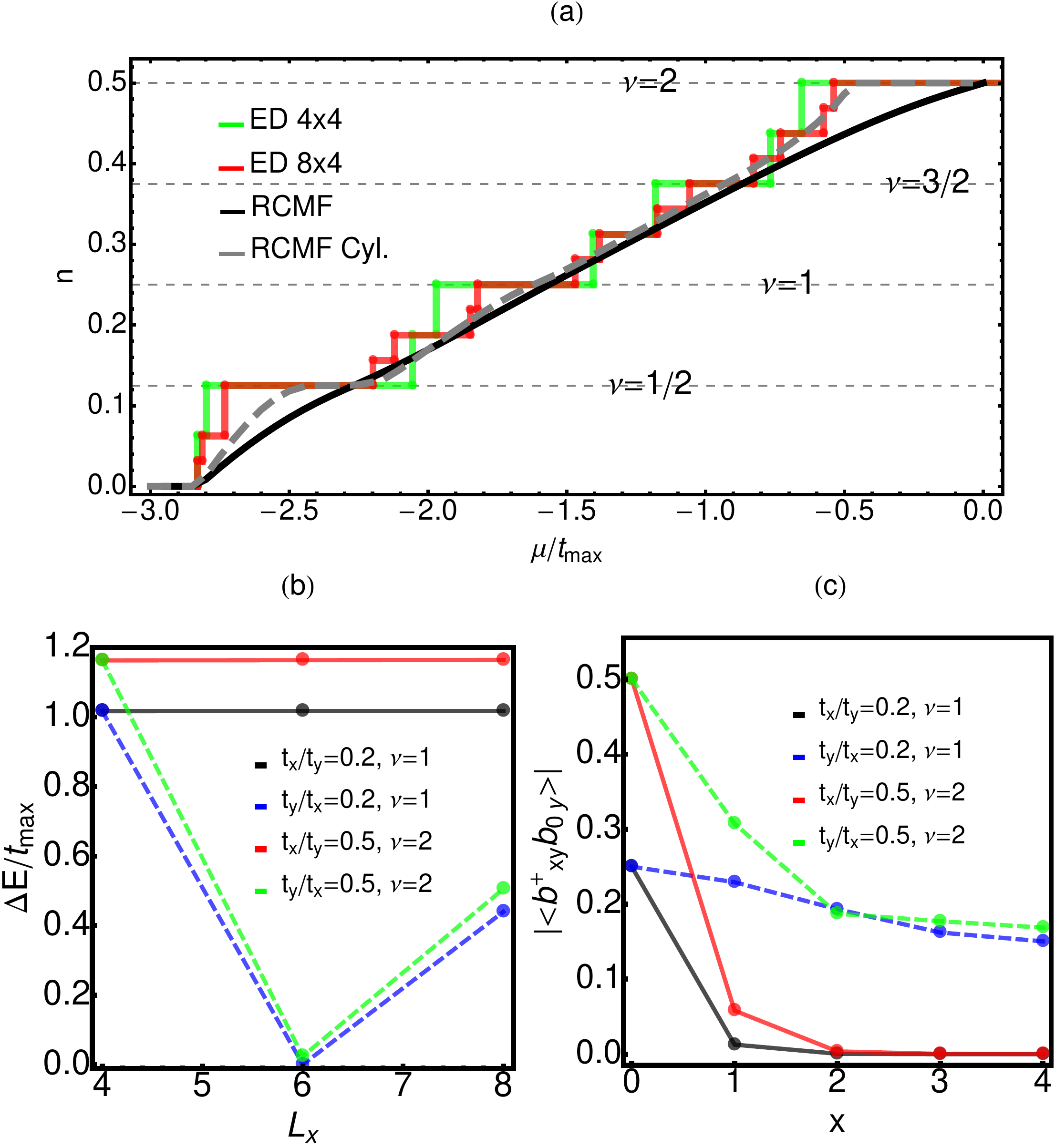}
  \caption{\label{fig:ED_comp} (a) Density $n$ as a function of $\mu$ at  $t_x=t_y$ computed with ED on a $4\times 4$ (green) and $8\times 4$ (red) system, compared to RCMF results on the square lattice (``RCMF", black) and on a cylinder with $4$ sites and periodic boundaries in the $y$-direction (``RCMF Cyl.", gray dashed). (b) Manybody gap for periodic boundaries and $L_y=4$ as a function of $L_x$ for $\nu=1$ and $t_x/t_y=0.2$ (black), $t_y/t_x=0.2$ (blue, dashed), and for $\nu=2$ and  $t_x/t_y=0.5$ (red), $t_y/t_x=0.5$ (green, dashed). (c) Correlations $\left|\langle b^{\dagger}_{x,y} b_{0,y}\rangle\right|$ as a function of x on a system with periodic boundaries, $L_x=8$, and $L_y=4$ for the same values and colors as in (b). } 
\end{figure}

\section{Charge conjugation relations of hard-core bosons}
\label{cchb}

For hard-core bosons a particle-hole transform (i.e.\ simultaneous $b^{\dagger}\rightarrow b$ and $b\rightarrow b^{\dagger}$) is equivalent to an inversion of the flux in the Hamiltonian of Eq.\ (\ref{eq:HofHarp}), i.e.\ $\Phi\rightarrow -\Phi$. A direct consequence of this, is that the manybody groundstates at densities $n$ and $1-n$ are related by the $\mathcal{C}\mathcal{T}$ operation, where $\mathcal{C}$ is the particle-hole transform, and  $\mathcal{T}$ the complex conjugation operator. This implies that the Hall conductivity $\sigma_{xy}$ is anti-symmetric under the transform $n\rightarrow 1-n$ \cite{Lind_10}, i.e.
\begin{equation}
\sigma_{xy}(n)=-\sigma_{xy}(1-n).
\label{Hall}
\end{equation}
This effect is known as the charge conjugation symmetry of hard-core bosons \cite{Lind_10}.

Further, the uncondensed phase at $\nu=2$ ($n=1/2$) discussed in Sec.\ \ref{incom} is by definition $\mathcal{C} \mathcal{T}$-symmetric with zero Hall conductivity. This implies that in the chiral current patterns of Fig.\ \ref{fig:iQHE}b, the ``charge" transport of the ``particle" and ``hole" channels will always cancel each other. 

\section{Comparison with exact diagonalisation}
\label{Comp_ED}

In Fig.\ \ref{fig:ED_comp}a we compare our RCMF results with ED results on finite systems. The $4\times 4$ ED system differs from the $F=0$ solution of RCMF by a renormalization of the hopping according to Eq.\ (\ref{renor_hop}), resulting in a shift in chemical potential of the plateaus. We present a sweep of the density in chemical potential without hopping anisotropy (i.e. $t_x=t_y$). As can be seen the only regions where we see a large discrepancy with respect to ED are around fillings $\nu=1/2$ and $\nu=2$. These are the fillings where the metastable plateaus are particularly close in energy to the symmetry-broken groundstate (see Fig.\ \ref{fig:EnDif}). 

We further compare the ED results with RCMF results on a cylinder with just $4$ sites and periodic boundary conditions in $y$-direction. This can easily been done by modyfying the coarse-graining procedure of Eq.\ (\ref{eq:epsilon_bar}), which is now only integrated over $k$. This results in a new cluster-hopping
\begin{align}
\bar{t}_{\text{\tiny $(X\pm 1,Y)$,$(X,Y)$}} &=\frac{2\sqrt{2}}{\pi}t_{\text{\tiny $(X \pm 1,Y)$,$(X,Y)$}}, \nonumber \\
\bar{t}_{\text{\tiny $(X,Y\pm 1)$,$(X,Y)$}} &=t_{\text{\tiny $(X,Y\pm 1)$,$(X,Y)$}}. \nonumber
\end{align}
As can be seen in  Fig.\ \ref{fig:ED_comp}a, in this case also RCMF shows a fractional plateau at $\nu=1/2$ and a plateau at $\nu=2$, indicating that at zero anisotropy these phases are much more robust in the cylinder geometry than they are on the infinite square lattice. 

Apart from the response to twisted boundaries discussed in Sec.\ \ref{incom}, the anisotropic gapless nature of the two-dimensional uncondensed phases can also be observed in the scaling of the many-body gap as a function  of $L_x$ while keeping $L_y=4$ fixed. As shown in Fig. \ref{fig:ED_comp}b the manybody gaps remain essentially constant if $t_x$ is (sufficiently) smaller than $t_y$, while it decreases in a non-monotonous way if $t_x$ is larger than $t_y$. If the same scaling is done in $y$-direction the situation is reversed. This also implies that on the cylinder these phases are gapped for $t_x<t_y$. The same behavior can also be observed in the correlations $|\langle b^{\dagger}_{x,y} b_{0,y}\rangle|$ in a system with $L_y=4$ and $L_x=8$, shown in Fig.\ \ref{fig:ED_comp}c, which quickly drop to zero as a function of $x$ for $t_x<t_y$, indicating a gapped phase on the cylinder. For $t_x>t_y$ on the other hand, the correlations stay finite throughout the whole system hinting at the anisotropic gapless nature of the phase in two dimensions (and on the cylinder if $t_x$ is sufficiently large).

\section{Topological properties}
\label{Chern}

\begin{figure}
  \includegraphics[scale=.38]{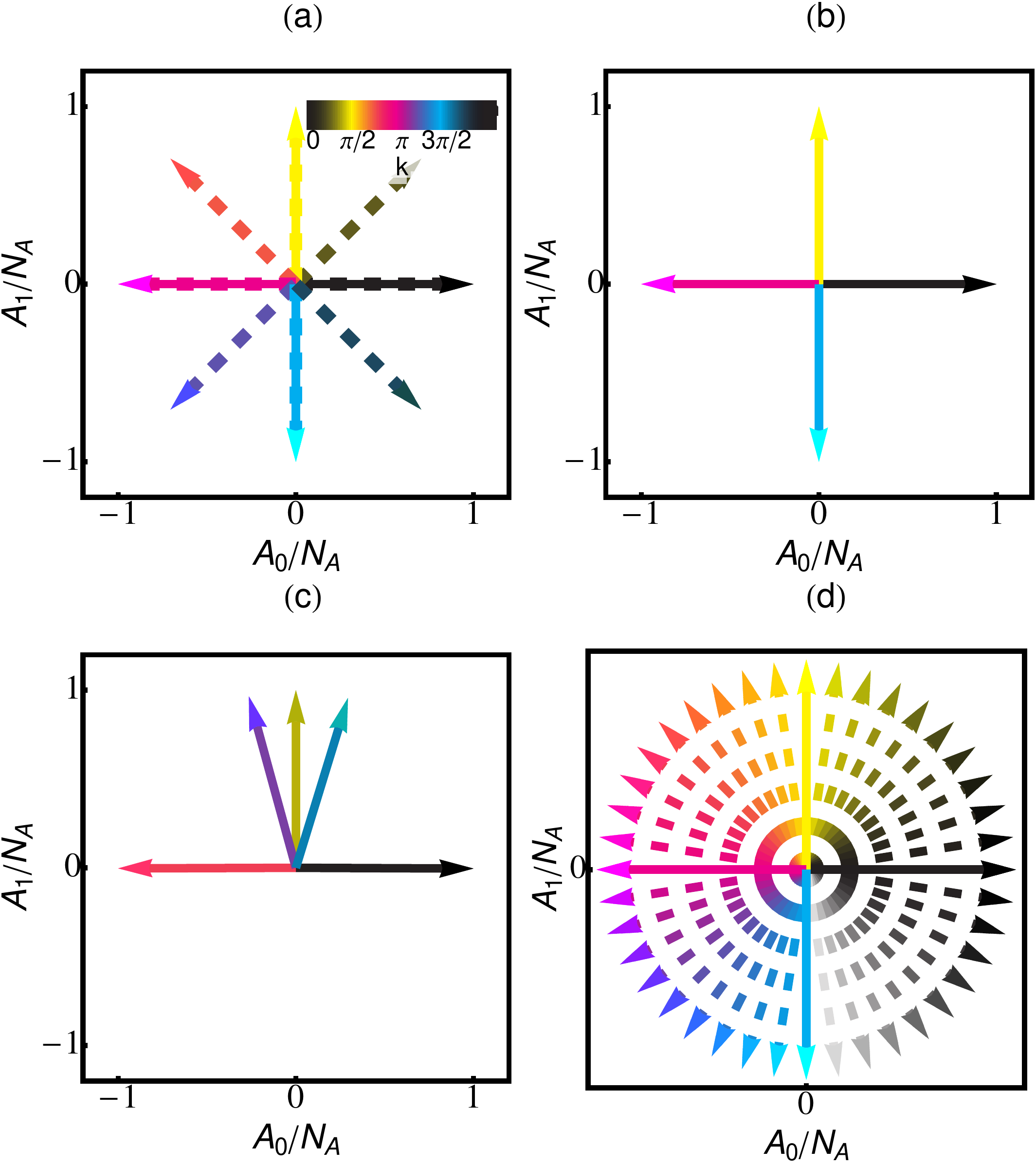}
  \caption{\label{fig:ED_cher} (a) Values of the $\vec{h}$ vector as a function of momentum for $t_x/t_y=0.5$ and $\nu=2$, on a $4\times 4$ (full lines) and $8\times 4$ (dashed lines) system. (b) Values of the $\vec{h}$ vector as a function of momentum for $t_y/t_x=0.2$ and $\nu=1/2$, on a $4\times 4$ system. (c) Values of the $\vec{h}$ vector as a function of momentum for $t_y/t_x=0.2$ and $\nu=3/5$ on a $5\times 4$ system. (d) Response of the momentum-values of $\vec{h}_{k,q}$ to the boundary twisting angle $\theta_x\in \left[0,2\pi\right]$ on a cylinder geometry ($\theta_y=0$) for $t_x/t_y=0.2$ and $\nu=1$ on a $4\times 4$ system, where the dashed lines indicate $\theta_x\neq 0$, while the coloring indicates the value of $k+\theta_x/L_x$.} 
\end{figure}
Since RCMF does not give direct access to the many-body groundstate of the infinite lattice, nor to dynamical quantities, there is no way to directly compute the many-body Chern number of the system. Instead, we make use of the properties of the lattice to indirectly measure the topology of the groundstate using the $\hat{h}_{k,q}$ vector introduced in Sec.\ \ref{Sec_Mod}.
It should be noted that the winding of $\hat{h}_{k,q}$ is independent of the basis, since the geometric angle of a vector $\vec{u}(x,y)$ remains the same under an axis rotation $y\rightarrow y'=y\cos{\alpha}$, if $\alpha$ is not an odd multiple of $\pi/2$. 
\begin{figure}
  \includegraphics[scale=.85]{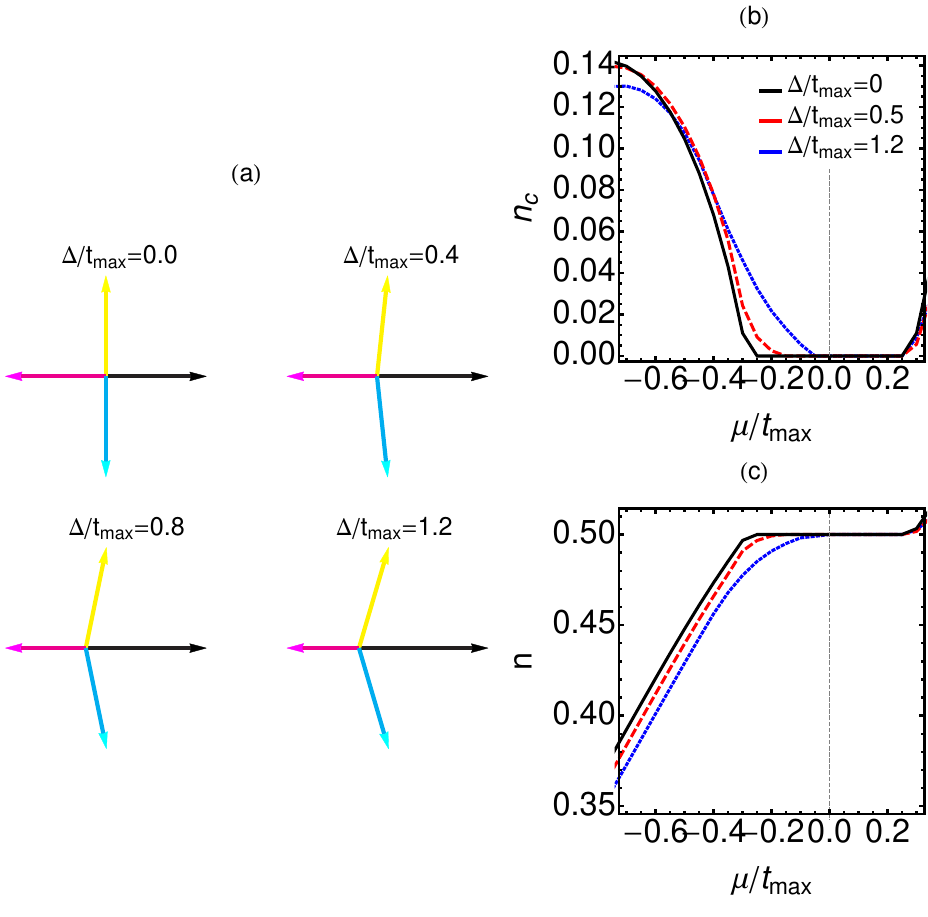}
  \caption{\label{Fig_Dis} Stability of the winding against local perturbations. (a) Winding of the $\vec{h}_{k,q}$-vector for $\nu=2$ at $t_x/t_y=0.4$ and $\mu=0$ for different perturbation strengths $\Delta$ [see Eq.\ (\ref{pertur})]. (b) and (c) Sweeps in chemical potential of the condensate density $n_c$ and the total density $n$ at $(t_x-t_y)/t_{\rm max}=-0.6$ and $\Delta/t_{\rm max}=0$ (black), $\Delta/t_{\rm max}=0.5$ (red dashed), and $\Delta/t_{\rm max}=1.2$ (blue).}
\end{figure}

In our RCMF approach Eq.\ (\ref{eq:Omega1}) reduces to $\Omega=\Omega'$ in the absence of $U(1)$ symmetry-breaking. Computing $\hat{h}_{k,q}$ in the phases with $F=0$ by taking expectation values for the discrete momentum values of the cluster ($K$ and $Q$) then is equivalent to taking the same expectation values with respect to the infinite lattice. By looking at the values of $\hat{h}_{K,Q}$ at these discrete momenta and extrapolating its rotation on  the infinite lattice, we are thereby able to measure the topology of the infinite lattice in a way that is not limited by finite-size effects. This is shown in Fig.\ \ref{fig:ED_cher}a, where the $\hat{h}_{k,q}$ vector is compared in a $4\times 4$ and $8\times 4$ periodic system, respectively, for filling $\nu=2$ and $(t_x-t_y)/t_{\rm max}=-0.5$, yielding excellent agreement. In Figs.\ \ref{fig:ED_cher}b and \ref{fig:ED_cher}c we show the precession of $\hat{h}_{k,q}$ for filling $\nu=1/2$ on a $4\times 4$ system, and  $\nu=3/5$ on a $5\times 4$ system. At $\nu=1/2$ the system is in a fQH phase showing a topological winding. At $\nu=3/5$, where for bosons no fQH phase is possible, the vector does not show any closed loop and has a net geometric angle of zero.

By using twisted boundaries in $x$-direction (i.e. varying $\theta_x$), while keeping $\theta_y=0$, we measure the response of the vector $\hat{h}_{k,q}$ on a cylinder to a magnetic flux piercing the system in $y$-direction in Fig.\ \ref{fig:ED_cher}d. As discussed in Sec.\ \ref{sec:nonint}, under the insertion of a flux of $\Phi_{\rm y}=\theta_x=2\pi$, the winding of the vector indicates an adiabatic translation of the manybody groundstate by one site in $y$-direction. For $\nu=1$ (shown in Fig.\ \ref{fig:ED_cher}d) this translates into a quantized transverse transport of a single particle around the periodic boundaries in $y$-direction, while for $\nu=3$ a single hole is being transported. In the case of $\nu=2$ the total charge transport is zero, as a particle-hole pair is transported.

We further analyze the stability of the winding number against local perturbations. To that end we introduce a local shift in chemical potential $\Delta$, such that the chemical potential is shifted on site $(X,Y)=(0,0)$ on the cluster, i.e.
\begin{equation}
\mu_{\text{\tiny X,Y}}=\mu-\Delta\delta_{\text{\tiny X,$0$}}\delta_{\text{\tiny Y,$0$}}.
\label{pertur}
\end{equation}
In Fig.\ \ref{Fig_Dis} we show the response of the uncondensed phases on a $4\times 4$ system at $\nu=2$ to this local perturbation. As can be seen in Fig.\ \ref{Fig_Dis}a, while the individual momentum values of the $\vec{h}_{k,q}$-vector change with the strength of the perturbation $\Delta$, the total winding of the vector around the origin remains stable even at such large values as $\Delta\approx t_{\rm max}$, while the single-particle gap (i.e. the size of the plateau) decreases (see Figs.\ \ref{Fig_Dis}b and  \ref{Fig_Dis}c). At larger values of $\Delta$ the phase has a non-zero condensate order parameter. A similar behavior can also be observed in the $\nu=1,3$ phases. In the full two-dimensional system we expect this robustness to vanish as the system-size is increased and the manybody gap goes to zero. In the cylindrical geometry however this points to a stability of the winding against disorder, indicative of topological protection.

\section{Mean-field decoupling in reciprocal space}
\label{Decoup}

In this section we give the details of the mean-field decoupling in reciprocal space within RCMF. We start from the separation of the hopping Hamiltonian $H_0$ [Eq.\ (\ref{hop_ham})] into a cluster-local part $H_{c}$ and an inter-cluster part $\Delta H$, i.e. $H_{0}=H_{\rm c}+\Delta H$ (see Sec.\ \ref{Sec_RCMF}). We now proceed with decoupling the inter-cluster part 
\begin{equation}
\Delta H = \sum_{K,Q}\sum_{\tilde{x},\tilde{y}}\sum_{\tilde{x}',\tilde{y}'}\delta {\epsilon}_{\text{\tiny K,Q}}(\tilde{x}-\tilde{x}',\tilde{y}-\tilde{y}')b^{\dagger}_{\text{\tiny K,Q}}(\tilde{x},\tilde{y})b_{\text{\tiny K,Q}}(\tilde{x}',\tilde{y}') ,
\nonumber
\end{equation}
through the decomposition of the creation/annihilation operators into their static expectation values $\phi_{\text{\tiny K,Q}}(\tilde{x},\tilde{y})=\langle b_{\text{\tiny K,Q}}(\tilde{x},\tilde{y})\rangle$ and fluctuations $\delta b$, i.e.
\begin{equation}
b_{\text{\tiny K,Q}}(\tilde{x},\tilde{y})=\phi_{\text{\tiny K,Q}}(\tilde{x},\tilde{y})+\delta b_{\text{\tiny K,Q}}(\tilde{x},\tilde{y}),
\label{decoup1}
\end{equation}

This approach decomposes $\Delta H$ into three separate parts
\begin{equation}
\Delta H=\Delta H_{\phi}+H_{\phi}+H_{\delta},\nonumber
\end{equation}
where $\Delta H_{\phi}$ is linear in $b$, and $b^{\dagger}$,
\begin{multline}
\Delta H_{\phi} = \sum_{K,Q}\sum_{\tilde{x},\tilde{y}}\sum_{\tilde{x}',\tilde{y}'}\delta{\epsilon}_{\text{\tiny K,Q}}(\tilde{x}-\tilde{x}',\tilde{y}-\tilde{y}') \\
\times \left(b^{\dagger}_{\text{\tiny K,Q}}(\tilde{x},\tilde{y})\phi_{\text{\tiny K,Q}}(\tilde{x}',\tilde{y}')+\phi^{*}_{\text{\tiny K,Q}}(\tilde{x},\tilde{y})b_{\text{\tiny K,Q}}(\tilde{x}',\tilde{y}')\right),\nonumber
\end{multline}
$H_{\phi}$ is the constant contribution
\begin{equation}
H_{\phi} = -\sum_{K,Q}\sum_{\tilde{x},\tilde{y}}\sum_{\tilde{x}',\tilde{y}'}\delta{\epsilon}_{\text{\tiny K,Q}}(\tilde{x}-\tilde{x}',\tilde{y}-\tilde{y}')\phi^{*}_{\text{\tiny K,Q}}(\tilde{x},\tilde{y})\phi_{\text{\tiny K,Q}}(\tilde{x}',\tilde{y}'),\nonumber
\end{equation}
and $H_{\delta}$ contains all quadratic fluctuations 
\begin{equation}
H_{\delta} = \sum_{K,Q}\sum_{\tilde{x},\tilde{y}}\sum_{\tilde{x}',\tilde{y}'}\delta{\epsilon}_{\text{\tiny K,Q}}(\tilde{x}-\tilde{x}',\tilde{y}-\tilde{y}')\delta b^{\dagger}_{\text{\tiny K,Q}}(\tilde{x},\tilde{y})\delta b_{\text{\tiny K,Q}}(\tilde{x}',\tilde{y}').\nonumber
\end{equation}

The standard procedure of the mean-field decoupling approximation consists in neglecting quadratic fluctuations, i.e. $H_{\delta} \approx 0$.
Furthermore, we assume translational invariance between the different clusters
\begin{equation}
\phi_{\text{\tiny K,Q}}(\tilde{x},\tilde{y})=\phi_{\text{\tiny K,Q}}.
\label{Cond_GS}
\end{equation}
By $\sum_{\tilde{x},\tilde{y}}\delta{\epsilon}_{\text{\tiny K,Q}}(\tilde{x},\tilde{y})=\delta\epsilon_{\text{\tiny K, $0$,Q, $0$}}$, this reduces the cluster-coupling part of the Hamiltonian to
\begin{align}
\Delta H \approx & \sum_{\tilde{x},\tilde{y}}\left(\Delta H_{\tilde{x},\tilde{y}}+C_{\phi}\right),\nonumber \\
\Delta H_{\tilde{x},\tilde{y}}  =& \sum_{K,Q}\delta\epsilon_{\text{\tiny K, $0$,Q, $0$}} 
\left( b_{\text{\tiny $K$,$Q$}}^{\dagger}(\tilde{x},\tilde{y}) \phi_{\text{\tiny $K$,$Q$}}
+\phi^{*}_{\text{\tiny $K$,$Q$}} b_{\text{\tiny $K$,$Q$}}^{}(\tilde{x},\tilde{y}) \right),\nonumber
\label{eq:DeltaH_mf}
\end{align}
with a constant scalar shift $C_{\phi}$, which for simplicity in the following will be omitted in the Hamiltonian (but has to be taken into account for the free energy), given by
\begin{equation}
C_{\phi}=-\sum_{K,Q}\delta\epsilon_{\text{\tiny K, $0$,Q, $0$}}\left|\phi_{\text{\tiny $K$,$Q$}}\right|^{2}.
\label{cphi}
\end{equation}

The system now consists of $\left(NM\right)/\left(N_c M_c\right)$ identical decoupled clusters with individual Hamiltonians
\begin{equation}
H_{\tilde{x},\tilde{y}}=\sum_{K,Q}\bar{\epsilon}_{\text{\tiny K,Q}}b^{\dagger}_{\text{\tiny K,Q}}(\tilde{x},\tilde{y})b_{\text{\tiny K,Q}}(\tilde{x},\tilde{y})+\Delta H_{\tilde{x},\tilde{y}}, \nonumber
\end{equation}
which, after a Fourier transform to position space, and dropping the $(\tilde{x},\tilde{y})$-notation, yields the effective mean-field Hamiltonian
\begin{align}
H_{\rm eff}=&\sum_{X',Y'}\sum_{X,Y}\bar{t}_{\text{\tiny $(X',Y')$,$(X,Y)$}}b^{\dagger}_{\text{\tiny $X'$, $Y'$}}b^{}_{\text{\tiny $X$, $Y$}}\nonumber\\
&+\sum_{X,Y}\left(b^{\dagger}_{\text{\tiny $X$, $Y$}}F_{\text{\tiny $X$, $Y$}}+F^{*}_{\text{\tiny $X$, $Y$}}b^{}_{\text{\tiny $X$, $Y$}}\right),\nonumber
\label{eq:H_mf}
\end{align}
where the symmetry breaking field $F_{\text{\tiny $X$, $Y$}}$ is given by
\begin{equation}
F_{\text{\tiny $X$, $Y$}}=\sum_{X',Y'}\delta t_{\text{\tiny $(X,Y)$,$(X',Y')$}}\phi^{}_{\text{\tiny $X'$, $Y'$}} \label{eq:symm1}
\end{equation}
and
\begin{align}
& \bar{t}_{\text{\tiny $(X',Y')$,$(X,Y)$}}=\frac{1}{N_c M_c}\sum_{K,Q}e^{\text{\tiny $i\left(K\left(X'-X\right)+Q\left(Y'-Y\right)\right)$}}\bar{\epsilon}_{\text{\tiny $K,Q$}}, \nonumber\\
& \delta t_{\text{\tiny $(X',Y')$,$(X,Y)$}}=t_{\text{\tiny $(X',Y')$,$(X,Y)$}}-\bar{t}_{\text{\tiny $(X',Y')$,$(X,Y)$}}.  \label{eq:delta_t1}
\end{align}

If instead of a pure hopping Hamiltonian, the Hamiltonian also includes local (interaction) terms, e.g.
\begin{equation}
H^{'}=H_{0}+H_{\rm int}=H_{0}+\frac{U}{2}\sum_{x,y} n_{\text{\tiny $x,y$}}\left(n_{\text{\tiny $x,y$}}-1\right)-\mu\sum_{x,y}n_{\text{\tiny $x,y$}},\nonumber
\end{equation}
the local part $H_{\rm int}$ is already inherently cluster-local and can be absorbed into $H_{\rm c}$ in Eq.\ (\ref{eq:Sep_Ham}), such that the effective Hamiltonian becomes
\begin{equation}
H^{'}_{\rm eff}=H^{}_{\rm eff}+H_{\rm int}.
\label{eq:H1_mf}
\end{equation}

Taking into account the constant shift of Eq.\ (\ref{cphi}), the free-energy of the full lattice system under the mean-field decoupling approximation can now be expressed  as
\begin{equation}
\Omega = \Omega^{'}-\frac{1}{2}\sum_{X,Y}\left(\phi^{*}_{\text{\tiny $X$, $Y$}} F_{\text{\tiny $X$, $Y$}}+F^{*}_{\text{\tiny $X$, $Y$}}\phi^{}_{\text{\tiny $X$, $Y$}}\right),
\label{eq:Omega}
\end{equation}
where $\Omega^{'}$ is the free energy of the cluster with the Hamiltonian of Eq.\ (\ref{eq:H1_mf}). 
With Eqs.\ (\ref{eq:symm1})-(\ref{eq:Omega}) we have everything in place in order to formulate the full RCMF approach, see Eqs.\ (\ref{eq:H11_mf})-(\ref{eq:SelfCon}) in Sec.\ \ref{Sec_RCMF}.

\section{Benchmarking RCMF}
\label{DCAMF}

\begin{figure*}
  \includegraphics[scale=.55]{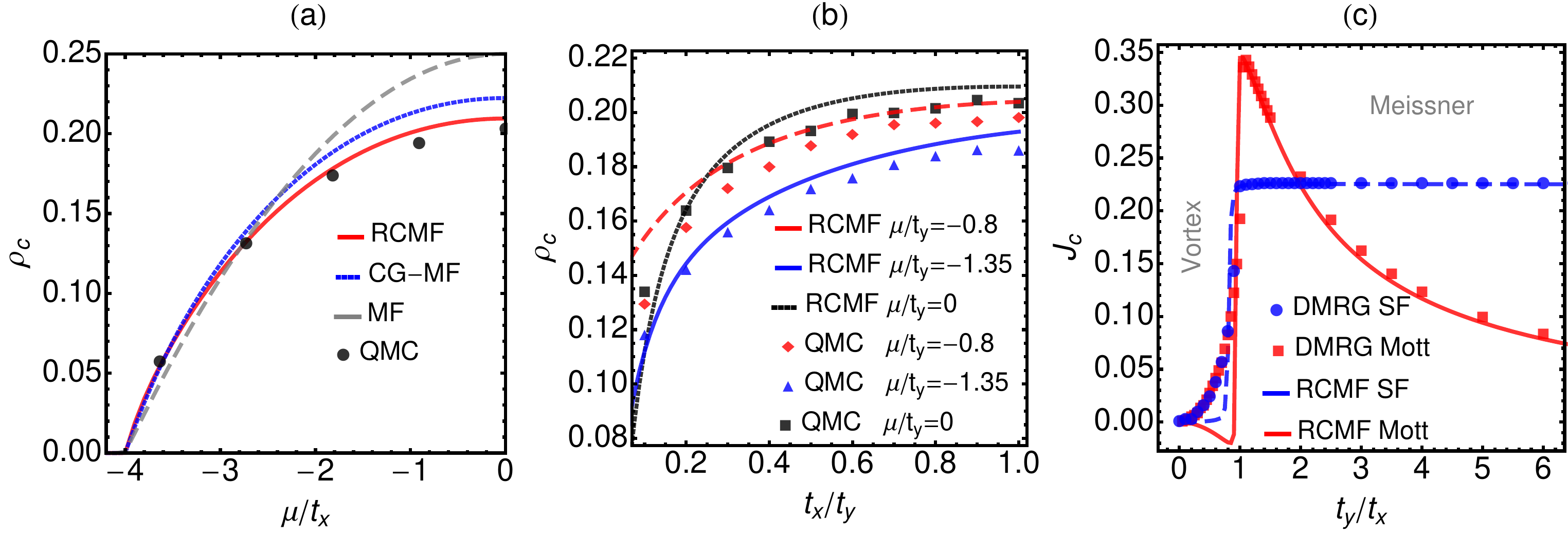}
  \caption{\label{fig:Bench} Benchmarking of RCMF. (a) Sweep of the condensate density $\rho_c$ in chemical potential $\mu$ for the Bose-Hubbard model with hard-core bosons on a 2d square lattice for $t_x=t_y=1$. The data are computed with QMC (black dots), RCMF on a $4\times 4$ cluster (red), CGMF on the same cluster (blue dotted) and standard single-site mean field (gray dashed). (b) Sweep of the condensate density $\rho_c$ of  the Bose-Hubbard model with hard-core bosons on a 2d square lattice as a function of $t_x/t_y$,  for fixed chemical potentials $\mu/t_y=0$ (black), $\mu/t_y=-0.8$ (red) and  $\mu/t_y=-1.35$ (blue). RCMF data are shown as lines, while QMC data are shown as dots. (c) Chiral current $J_{\rm c}$ [equation (\ref{eq:Jc})] of the chiral ladder of Refs. \onlinecite{Lad_Marie} and \onlinecite{Lad_Belen} with hard-core bosons for $\Phi=\pi/2$ as a function of anisotropy $t_x/t_y$. Results for $n=0.5$ (Mott) are shown in red, while results for $n=0.25$ (superfluid) are shown in blue. The RCMF results are shown as lines, while DMRG results  \cite{Lad_Marie} are shown as dots. } 
\end{figure*}
In order to benchmark RCMF we turn to the Bose-Hubbard model with hard-core bosons on a two-dimensional square lattice using a $4\times 4$ cluster Hamiltonian. In Fig.\ \ref{fig:Bench}a we show RCMF results for the condensate density $\rho_c=\sum_{X,Y}|\phi_{\text{\tiny $X,Y$}}|^2$ as a function of chemical potential for $t_x=t_y=1$ and compare with standard single-site mean field, CGMF \cite{Clust_MF} on a $4\times 4$ cluster, and numerically exact path integral quantum Monte Carlo (QMC) \cite{QMC_Cap,QMC_2D_ani} results. As expected, RCMF shows better agreement with QMC than the two other mean-field methods. In contrast to CGMF, which due to the breaking of translational invariance converges towards a weakly position-dependent (unphysical) condensate $\phi_{\text{\tiny $X,Y$}}$, the condensate in RCMF is completely homogeneous.

We also compare RCMF results with QMC for anisotropic systems in Fig. \ref{fig:Bench}b, observing stronger deviations with increasing anisotropy $|t_x-t_y|$. This is related to the use of a square symmetric $4\times 4$ cluster, while the bandwidths in $k$- and $q$-direction are no longer equal. As the one-dimensional limit ($t_x=0$) is approached, mean-field methods are always expected to behave worse, since quantum fluctuations play a bigger role. However, the results are still qualitatively correct, and we conclude that RCMF works reasonably well also for anisotropic systems.

In order to ensure that RCMF can properly treat artificial gauge fields, we simulate the two-leg ladder of Refs. \onlinecite{Lad_Belen} and \onlinecite{Lad_Marie} with a magnetic flux of $\Phi=\pi/2$ per plaquette and hard-core bosons using a $2\times 8$ cluster. This ladder corresponds to the Harper-Hofstadter-Mott model where the $x$-direction is restricted to just two sites. It shows Mott phases at density $n=0.5$ and superfluid phases otherwise, with both phases exhibiting Meissner and vortex current-patterns depending on the anisotropy \cite{Lad_Marie}. The Meissner phases can be found for anisotropies where for the gauge of Ref. \onlinecite{Lad_Belen} the non-interacting groundstate momenta -- i.e. the momenta where the dispersion has (degenerate) global minima -- are $k_{\rm gs}=\pm\pi/4$. These momenta are fully captured by the $2\times 8$ cluster with cluster-momenta $K=m\pi/4$, where $m=0,1,2,...,7$. On the other hand, in the anisotropy-region where the vortex phases appear, $k_{gs}$ varies as a function of the hopping-anisotropy \cite{Lad_Belen} and can no longer be represented within a $2\times 8$ cluster. This is shown in Fig.\ \ref{fig:Bench}c, where the RCMF chiral current
\begin{equation}
J_{\rm c}= \frac{1}{N}\sum_{y}\left(J_{\rm y}(0,y)-J_{\rm y}(1,y)\right)
\label{eq:Jc}
\end{equation}
is compared to DMRG results \cite{Lad_Marie} both in the Mott ($n=0.5$) and superfluid ($n=0.25$) regime. Here, $J_{\rm y}(l,y)$ is the current in $y$-direction on the $y$th site of the ladder-leg $l$. The RCMF results agree very well in the Meissner phases, while they cannot capture the vortex phases. This is a good example of what RCMF can do and what not: for RCMF to work it is indispensable that the cluster is both an integer multiple of the unit cell and that the groundstate momenta of the non-interacting model can be reproduced exactly by the grid of cluster momenta spanned by $K$ and $Q$. This is a direct consequence of Eq.\ (\ref{Cond_GS}), which in reciprocal space implies $\phi_{\text{\tiny K$+\tilde{k}$,Q$+\tilde{k}$}}=\phi_{\text{\tiny K,Q}}\delta_{\tilde{k},0}\delta_{\tilde{q},0}$. This condition, which was used to decouple the clusters in RCMF, can only be fulfilled if the momenta of the minima of the dispersion (which are the momenta where condensation will occur for local single-site interactions) can be reproduced by the momenta of the cluster $(K,Q)$. If this is the case, as seen in Fig. \ref{fig:Bench}c, the deviation from the DMRG results on  the chiral current \cite{Lad_Marie} is below $1\%$. 

\section{RCMF approach for the Harper-Hofstadter-Mott model}
\label{hhmm}

The HHm has groundstate momenta $k_{\rm gs}=0,\pm\pi/2,\pi$ and $q_{\rm gs}=0$. Since the momenta of the groundstate are independent of the anisotropy, we do not encounter the difficulties described in Appendix \ref{DCAMF} for the vortex phases of the chiral ladder when using finite clusters. In order to fulfill Eq.\ (\ref{Cond_GS}) by reproducing the minima of the dispersion (see Appendix \ref{DCAMF}) a multiple of $4$ sites in $X$ direction is needed, since for $4$ sites $K$ is a multiple of $\pi/2$. We also need a multiple of $4$ sites in $Y$ direction in order to fully capture the $1\times 4$ unit cell. In this work we restrict ourselves to the minimal cluster, i.e.  $4\times 4$. 

Since the mean-field decoupling is performed in the thermodynamic limit, the sum over $\tilde{k}$ and $\tilde{q}$ in (\ref{eq:epsilon_bar}) can be replaced by an integral and computed analytically. In this configuration the coarse-graining described in Sec.\ \ref{Sec_RCMF} leads to the cluster-hopping
\begin{equation}
\bar{t}_{\text{\tiny $(X',Y')$,$(X,Y)$}}=\frac{2\sqrt{2}}{\pi}t_{\text{\tiny $(X',Y')$,$(X,Y)$}},
\label{renor_hop}
\end{equation}
where $t_{\text{\tiny $(X',Y')$,$(X,Y)$}}$ is the original hopping of the HHm [Eq.\ (\ref{eq:HofHarp_xy})] with periodic boundary conditions, which plugged into (\ref{eq:H11_mf}-\ref{eq:delta_t}) yields the effective RCMF Hamiltonian for the Harper-Hofstadter-Mott model on a $4\times 4$ cluster. 

\subsection{Observables}

The free energy $\Omega$ of Eq.\ (\ref{eq:Omega1}) represents the free energy of the lattice system in the thermodynamic limit within the RCMF approximation.
Using functional derivatives of Eq.\ (\ref{eq:Omega1}) we can compute expectation values with respect to the full lattice system. According to the self-consistency condition [Eq.\ (\ref{eq:SelfCon})], this is trivial for the condensate
\begin{equation}
\phi^{}_{\text{\tiny $X$, $Y$}}=\langle b_{\text{\tiny $X$, $Y$}}\rangle.\nonumber
\end{equation}
Accordingly, we get for the condensate density
\begin{equation}
\rho_c(X,Y)=\left|\phi^{}_{\text{\tiny $X$, $Y$}}\right|^2,\nonumber
\end{equation}
and the total condensate density per site
\begin{equation}
n_c=\frac{1}{N_c M_c}\sum_{X,Y}\rho_c(X,Y).\nonumber
\end{equation}

Also for the particle density we get an equivalence between the full lattice and the $4\times 4$ cluster, since
\begin{equation}
\rho(X,Y)=-\frac{\delta\Omega}{\delta\mu_{\text{\tiny X,Y}}}=\langle n_{\text{\tiny $X$, $Y$}}\rangle,\nonumber
\end{equation}
and, accordingly, for the total particle density per site
\begin{equation}
n=\frac{1}{N_c M_c}\sum_{X,Y}\rho(X,Y).\nonumber
\end{equation}

The current $J_{\rm x}(x,y)$ in $x$-direction between the sites $(x,y)$ and $(x+1,y)$ is defined as
\begin{multline}
J_{\rm x}(x,y)=-i\left(t_{\text{\tiny $(x+1,y),(x,y)$}}\langle b^{\dagger}_{\text{\tiny $x+1,y$}}b^{}_{\text{\tiny $x,y$}}\rangle_{\rm latt}\right. \\
\left. -t_{\text{\tiny $(x,y),(x+1,y)$}}\langle b^{\dagger}_{\text{\tiny $x,y$}}b^{}_{\text{\tiny $x+1,y$}}\rangle_{\rm latt}\right),\label{eq:Jx}
\end{multline}
where $\langle \text{.} \rangle_{\rm latt}$ is the lattice-system expectation value. However, by
\begin{multline}
\langle b^{\dagger}_{\text{\tiny $x',y'$}} b^{}_{\text{\tiny $x,y$}} \rangle_{\rm latt}  = \frac{\partial \Omega}{\partial t_{\text{\tiny $(x',y'),(x,y)$}}} = \frac{\partial\bar{t}_{\text{\tiny $(x',y'),(x,y)$}}}{\partial t_{\text{\tiny $(x',y'),(x,y)$}}}\langle b^{\dagger}_{\text{\tiny $x',y'$}}  b^{}_{\text{\tiny $x,y$}} \rangle \\
+\frac{1}{2}\frac{\partial\delta{t}_{\text{\tiny $(x',y'),(x,y)$}}}{\partial t_{\text{\tiny $(x',y'),(x,y)$}}}\left(\langle b^{\dagger}_{\text{\tiny $x',y'$}} \rangle \phi^{}_{\text{\tiny $x,y$}} +\phi^{*}_{\text{\tiny $x',y'$}} \langle b^{}_{\text{\tiny $x,y$}} \rangle\right), 
\label{eq:exp_vals}
\end{multline}
we can express the lattice quantities in terms of expectation values with respect to the RCMF Hamiltonian.

By Eq.\ (\ref{eq:exp_vals}) we can also compute the current in the $y$-direction given by
\begin{multline}
J_{\rm y}(x,y)=-i\left(t_{\text{\tiny $(x,y+1),(x,y)$}}\langle b^{\dagger}_{\text{\tiny $x,y+1$}}b^{}_{\text{\tiny $x,y$}}\rangle_{\rm latt}\right. \\
\left. -t_{\text{\tiny $(x,y),(x,y+1)$}}\langle b^{\dagger}_{\text{\tiny $x,y$}}b^{}_{\text{\tiny $x,y+1$}}\rangle_{\rm latt}\right),\label{eq:Jy}
\end{multline}
and by Fourier transform also the occupation in momentum space
\begin{equation}
\langle n_{\text{\tiny $K$, $Q$}}\rangle_{\rm latt}=\sum_{X,Y}e^{\text{\tiny $i\left(K X+Q Y\right)$}}\langle b^{\dagger}_{\text{\tiny $X,Y$}}b^{}_{\text{\tiny $0,0$}}\rangle_{\rm latt}.\nonumber
\label{eq:n_k}
\end{equation}

\subsection{Current-current correlations}
\label{Sec:cur_cur}

In the uncondensed phases described in Sec.\ \ref{Sec_Res} the current within the system is zero. However, as the system is not band insulating and the kinetic energy is non-zero this must result from two counter-propagating modes which cancel each other out. We can analyze these modes by measuring current-current correlations between neighboring bonds. By Eq.\ (\ref{eq:exp_vals}) we can write the currents on the full lattice $J_{\rm x/y}$ as expectation values of lattice-operators $\hat{J}_{\rm x/y}$ with respect to the RCMF Hamiltonian, i.e.
\begin{equation}
J_{\rm x/y}(x,y)=\langle \hat{J}_{\rm x/y}(x,y)\rangle
\nonumber
\end{equation}
with $\hat{J}_{\rm x/y}(x,y)$ defined according to Eqs.\ (\ref{eq:Jx})-(\ref{eq:Jy}). We now can look at the following current-current correlations:
\begin{align}
\Delta J_{{\rm x,x}}(x,y) &=\left\langle \hat{J}_{\rm x}(x,y)\hat{J}_{\rm x}(x+1,y)\right\rangle\nonumber\\
\Delta J_{{\rm x,y}}(x,y) &=\left\langle \hat{J}_{\rm x}(x,y)\hat{J}_{\rm y}(x,y)\right\rangle\nonumber\\
\Delta J_{{\rm x,y-1}}(x,y) &=\left\langle \hat{J}_{\rm x}(x,y)\hat{J}_{\rm y}(x,y-1)\right\rangle \nonumber
\end{align}

These three quantities are enough to describe the current patterns depicted in Fig.\ \ref{fig:iQHE}d: if $\Delta J_{{\rm x,x}}(x,y)$ is positive the currents $J_{\rm x}(x,y)$ and $J_{\rm x}(x+1,y)$ point in the same direction, if it is negative they point in opposite directions. The same is also true for $\Delta J_{{\rm x,y}}(x,y)$ and $\Delta J_{{\rm x,y-1}}(x,y)$. Assuming a finite current in $+x$ direction on a given site $(x,y)$ and extracting the sign of the currents on the neighboring bonds through the correlations introduced above, one can therefore easily draw one of the two counter-propagating current patterns. The other pattern results from simply inverting the direction of all currents.

\bibliography{HHHM1}

\end{document}